\documentstyle[preprint,aps,prd]{revtex}

\input{epsf}
\tightenlines

\begin{document}
\thispagestyle{empty}

\newcommand{\eq}{\begin{eqnarray}}
\newcommand{\en}{\end{eqnarray}}
\renewcommand{\theequation}{\arabic{section}.\arabic{equation}}

\newcommand{\bea}{\begin{eqnarray}}
\newcommand{\eea}{\end{eqnarray}}

\newcommand{\mpps}{M_{\pi^+}^2}
\newcommand{\mpzs}{M_{\pi^0}^2}

\newcommand{\mpp}{M_{\pi^+}}
\newcommand{\mpz}{M_{\pi^0}}

\newcommand{\scs}{\, , \,}
\newcommand{\per} {\, .}

\newcommand{\ra}{\rangle}
\newcommand{\la}{\langle}

\newcommand{\mathbold}[1]{\mbox{\boldmath $\bf#1$}}
\newcommand{\lagnr}{{\cal L}_{NR}}

\preprint{BUTP-01/03}

\draft
\title{Decays of the $\pi^+\pi^-$ atom}

\author{J. Gasser}
\address{Institute for Theoretical Physics, University of Bern,
         Sidlerstrasse 5, CH-3012, Bern, Switzerland}
\author{V. E. Lyubovitskij}
\address{Institute of Theoretical Physics, University of T\"{u}bingen, 
         Auf der Morgenstelle 14, D-72076 T\"{u}bingen, Germany, and
         Department of Physics, Tomsk State University,
         634050 Tomsk, Russia}
\author{A. Rusetsky}
\address{Institute for Theoretical Physics, University of Bern,
         Sidlerstrasse 5, CH-3012, Bern, Switzerland, and
         HEPI, Tbilisi State University, University St. 9,
         380086 Tbilisi, Georgia}
\author{A. Gall}
\address{SWITCH, Limmatquai 138, 8001 Z\"{u}rich}

\date{14 March 2001}

\maketitle

\begin{abstract}
We construct  an effective non-relativistic quantum field  theory 
that describes bound states of $\pi^+\pi^-$ pairs  and their
hadronic decays. We then derive a general expression 
for the lifetime  of the ground state  at next-to-leading order in
 isospin breaking.
 Chiral perturbation theory allows one to
 relate the decay rate to the two $S$-wave $\pi\pi$ scattering lengths and to 
 several low-energy constants that occur in the chiral Lagrangian.
Recent predictions for the scattering lengths
 give   $\tau=(2.9\pm 0.1)\times10^{-15}$ sec.
This result may be confronted with $\pi^+\pi^-$ lifetime measurements,
 like the one presently carried out at CERN.
\end{abstract}


\tableofcontents

\newpage

\setcounter{equation}{0}
\section{Introduction}

The DIRAC experiment at CERN~\cite{DIRAC} aims to measure the lifetime of 
the $\pi^+\pi^-$ atom (pionium) in its ground state with high
  precision. This atom 
decays predominantly into two neutral pions,
$\Gamma\simeq\Gamma_{2\pi^0}$. The latter decay rate is
proportional to the square of the difference $a_0-a_2$ of the strong 
$S$-wave $\pi\pi$ scattering lengths~\cite{Deser,Uretsky} with isospin
$I=0,2$. The measurement will therefore allow one to determine 
this difference,  which may then be confronted with the predicted value
$a_0-a_2=0.265\pm 0.004$~\cite{Colangelo}.
What makes this enterprise particularly exciting is the fact that one may 
 determine in this manner  the nature of spontaneous chiral 
symmetry breaking in QCD by experiment: Should it turn out that the 
predictions~\cite{Colangelo} are in conflict with the results of DIRAC, 
one would have to conclude~\cite{Stern} that spontaneous chiral symmetry 
breaking in QCD differs from the standard
picture~\cite{ChPT,ChPTlit1,ChPTlit2}. An analogous determination of
the nature of spontaneous chiral symmetry breaking may be performed
through an analysis of $K_{e_4}$ decays, which allows one
to measure the scattering length $a_0$ \cite{ke4}.

In order to determine the scattering lengths through
 a measurement of the pionium lifetime, the theoretical expression 
for the width must 
 be known with a precision that matches the accuracy of the 
lifetime measurement of DIRAC. In Ref.~\cite{Bern1} (see also~\cite{Zuoz}), 
we have presented a
compact expression for $\Gamma_{2\pi^0}$ in the framework of QCD 
(including photons) by use of effective field theory techniques.
The result obtained contains all terms at leading and
next-to-leading order in the isospin breaking parameters 
$\alpha\simeq 1/137$ and $(m_u-m_d)^2$. On the basis of this formula, 
a numerical analysis  was carried out in 
Ref.~\cite{Bern2} at order $e^2p^2$ in ChPT. The aim of the present paper 
is: i) to give a complete description of the theory of the
$\pi^+\pi^-$ atom decay, providing the details that were omitted 
in~\cite{Bern1,Bern2}, and ii) to update that numerical analysis by use
of the information recently obtained in \cite{Colangelo} on the
scattering lengths and on one of the low-energy constants.
 
We first briefly review previous work on the 
subject. Theoretical investigations of hadronic atoms and, in particular, 
of $\pi^+\pi^-$ decay, have been performed in several settings. Potential 
scattering theory in the framework of quantum mechanics has  been used 
in~\cite{Deser,Trueman,Rasche,Minkowski}, and methods of quantum field 
theory have been invoked as well 
\cite{Efimov,Pervushin,Volkov,Jentschura,Sazdjian,Atom}.      
In particular, in Refs.~\cite{Sazdjian}, the lifetime of the 
$\pi^+\pi^-$ atom was calculated by use of two-body wave 
equations of 3D-constraint field theory. In Refs.~\cite{Atom}, the $\pi^+\pi^-$ atom
decay was studied in a field-theoretical approach based on the 
Bethe-Salpeter equation. The results for the $\pi^+\pi^-$ atom lifetime
obtained with the two latter approaches contain the major
next-to-leading order terms in isospin breaking and agree both conceptually 
and numerically. However, in these  investigations the momentum dependence 
of the strong $\pi\pi$ scattering amplitude was neglected.

In several recent publications~\cite{Labelle,Kong,Holstein,Soto1,Soto2,Soto3}, the decay 
of $\pi^+\pi^-$ atoms has been  studied in the framework of a 
non-relativistic effective Lagrangian - a method originally proposed by 
Caswell and Lepage~\cite{Lepage} to investigate bound states in general. 
This method has proven to be far more efficient than conventional 
approaches based on relativistic bound-state equations. It allows one e.g. 
to go beyond the approximation used in~\cite{Sazdjian,Atom} for the 
scattering amplitudes. In our previous publications~\cite{Bern1,Bern2}, 
we have used the same method. We refer the reader to \cite{Bern2} for a 
comparison of the various results obtained in the effective framework.

We now describe the general features of the system that we are going to study.
The $\pi^+\pi^-$ atom is a highly non-relativistic, loosely bound system. 
The  pions are mainly bound by the 
 Coulomb force, and the atom decays 
predominantly through the strong interactions. The average momentum of the 
constituents in the CM frame is $\sim 0.5~{\rm MeV}$, and the Bohr 
radius of the bound state is $\sim 400~{\rm fm}$. The decay width of the
$\pi^+\pi^-$ atom $\sim 0.2~{\rm eV}$ is much smaller then
the binding energy $\sim 2\times 10^{3}~{\rm eV}$. For this reason,
a non-relativistic framework provides the most economical and 
powerful approach to the calculation of the characteristics of this sort 
of bound states. Since the strong interactions between pions at low energy 
can be described with ChPT, the theory of the $\pi^+\pi^-$ atom 
turns out to be  a merger of a non-relativistic approach with 
ChPT. Owing to the might of the non-relativistic approach which almost
trivializes the calculations in the bound-state sector, we
are able to  determine the first few coefficients in the 
chiral expansion of the  bound-state observables.

The paper is organized as follows. In Section~\ref{sec:eff-theory} we discuss
the foundations of the theory: the non-relativistic Lagrangian, 
Green functions, and matching to the relativistic amplitudes. Bound states
 are discussed in Section~\ref{sec:bound-states}. Using
 Feshbach's formalism~\cite{Feshbach}, we derive a master equation for 
the position of the  poles in the resolvent.  
In Section~\ref{sec:pionium} we derive - on the basis of the
master equation - a general expression 
for the decay width of the $\pi^+\pi^-$ atom in the ground state,
valid at next-to-leading order in isospin breaking. 
We then express this quantity, through the matching condition, 
 by the  relativistic  $\pi^+\pi^-\rightarrow\pi^0\pi^0$ scattering 
amplitude at threshold. A  numerical
analysis of the decay width at order $e^2p^2$ in ChPT is also carried out in
this section. 
Section~\ref{sec:conclusions} contains our conclusions. Background
material is  relegated to the Appendixes: 
 In Appendix~\ref{app:Lagrangian}, we discuss
the construction of a general non-relativistic Lagrangian with pions and
photons. The scattering sector of the non-relativistic theory is discussed in
Appendix~\ref{app:scattering}. In particular,  we argue that 
 the contributions of  transverse photons  to the
$\pi^+\pi^-\rightarrow\pi^0\pi^0$ scattering amplitude vanish at 
threshold at order $e^2$.
Therefore, these diagrams may be omitted in  matching
 the  relativistic and
non-relativistic amplitudes. Appendix~\ref{app:bound-states} deals with the
bound states in the non-relativistic theory: we show that, for a large
class of diagrams,
 transverse photons do not contribute to the decay width at
next-to-leading order in isospin breaking. On the basis of the results 
obtained in Appendixes~\ref{app:scattering} and \ref{app:bound-states}, we 
completely eliminate transverse photons from the theory. 
 In Appendix~\ref{app:threshold} we compare two different matching
 procedures.
Finally, in Appendix~\ref{app:mapping}, the 
$SU(3)\times SU(3)\rightarrow SU(2)\times SU(2)$ mapping of the
pertinent combination of electromagnetic low-energy constants in ChPT 
is provided.

\setcounter{equation}{0}
\section{The effective non-relativistic theory}
\label{sec:eff-theory}

In the framework of QCD (including photons), the  
energy levels and decay widths of pionium  
 are functions of the fine-structure constant $\alpha\simeq 1/137$, of the 
quark masses and of the
renormalization group invariant scale of QCD. In the following, we  
concentrate on the width $\Gamma$ of the ground state. 
It can be expanded in  powers
of $\alpha$ and of the quark mass difference $m_d-m_u$ (up to
logarithms).
 The leading and next-to-leading order terms in this
expansion  are due to
the decay into two neutral pions \cite{Deser},
\eq\label{eqleading}
\Gamma&=&\Gamma_{2\pi^0}+O(\delta^{5})\,,\nonumber\\
\Gamma_{2\pi^0}&=&\frac{2}{9}\alpha^3 p^\star(a_0-a_2)^2
+O(\delta^{9/2})\,
; \, \, p^\star=(\mpps-\mpzs-\frac{1}{4}\mpps\alpha^2)^{1/2}\, ,
\en
where $a_0$ and $a_2$ denote  the 
 $S$-wave $\pi\pi$ scattering lengths with isospin $I=0$ and $I=2$,
 respectively.
We count $\alpha$ and $(m_d-m_u)^2$ as small parameters of
order $\delta$. 
 The leading term in the  decay width
 is then of order $\delta^{7/2}$.
We describe in  the present article in detail the  
evaluation of  $\Gamma_{2\pi^0}$ up to and including 
terms of order $\delta^{9/2}$, providing details omitted 
in Refs.~\cite{Bern1,Bern2}.

\subsection{Non-relativistic Lagrangian}
The  method used in \cite{Bern1} for describing the decay of loosely 
bound states is an adaption of  the
procedure proposed by Caswell and Lepage some time ago \cite{Lepage}
for describing bound states in quantum field theories.
 In the present case, we
need to formulate a non-relativistic quantum field theory that describes
strong and electromagnetic interactions of pions in the very low-energy
region. The relevant Lagrangian is a rather voluminous object - indeed, it
contains an infinite number of terms. Fortunately, in the present case,
 only a small subset  of that Lagrangian is finally needed.

The mathematical problem to be solved  may be formulated as follows:
Construct a non-relativistic Lagrangian ${\cal L}_{NR}$ that 
contains all terms needed to
evaluate the decay width $\Gamma$ up to 
and including terms of order $\delta^{9/2}$.
We  relegate the construction of this object  to the
Appendixes~\ref{app:Lagrangian},\ref{app:scattering} and 
 \ref{app:bound-states}, because the
intermediate steps require lengthy calculations, whereas the final answer is 
amazingly simple. Indeed, as already mentioned in \cite{Bern1}, the following
Lagrangian achieves  the goal:
\eq\label{Lagr_full}
{\lagnr}&=&{\cal L}_0+{\cal L}_D+{\cal L}_C+{\cal L}_S\, ,
\nonumber\\[2mm]
{\cal L}_0&=& \sum_{i=\pm, 0}\,\pi_i^\dagger\biggl( i\partial_t-M_{\pi^i}+
\frac{\triangle}{2M_{\pi^i}}\biggr)\pi_i,\quad\quad
{\cal L}_D=\sum_{i=\pm, 0}\,\pi_i^\dagger
\biggl(\frac{\triangle^2}{8M_{\pi^i}^3}+\cdots\biggr)\pi_i,
\nonumber\\[2mm]
{\cal L}_C&=&-4\pi\alpha(\pi_-^\dagger\pi_-)\triangle^{-1}
(\pi_+^\dagger\pi_+),
\nonumber\\[2mm]
{\cal L}_S&=&c_1\pi_+^\dagger\pi_-^\dagger\pi_+\pi_-
+c_2[\pi_+^\dagger\pi_-^\dagger(\pi_0)^2+{\rm h.c.}]
+c_3\,(\pi_0^\dagger\pi_0)^2
\nonumber\\[2mm]
&+&c_4[\pi_+^\dagger\stackrel{\leftrightarrow}
{\triangle}\pi_-^\dagger(\pi_0)^2+\pi_+^\dagger\pi_-^\dagger\pi_0
\stackrel{\leftrightarrow}{\triangle}\pi_0+{\rm h.c.}],
\en
where  $(u\stackrel{\leftrightarrow}{\triangle}v)\doteq 
u \triangle v + v \triangle u$, and where $\triangle^{-1}$ denotes the inverse
of the Laplacian.

The Lagrangian contains explicitly only the pionic degrees of freedom - the
sole remnant of the photons is contained in the Coulomb interaction described
by ${\cal L}_C$. The mass parameters $M_{\pi^i}$ coincide with the physical masses
of the charged ($\mpp$) and neutral ($\mpz$) pions. The role of the low-energy
constants (LECs) $c_1,\ldots,c_4$ is discussed below.

\subsection{Green functions at $\alpha=0$}
\label{sec:alpha=0}
The fundamental objects  in the non-relativistic theory are
Green functions of the pion fields. They are most straightforwardly 
evaluated with path
integral techniques. For instance, the propagators of the free fields,
associated with ${\cal L}_0$, read
\eq\label{propagator_0}
{G}^0_{NR,i}(x)&=&{(2\pi)^{-4}}
\int \frac{d^4p \,\, e^{-ipx}}{M_{\pi^i}+{\bf p^2}/2M_{\pi^i}-p^0-i\epsilon}\nonumber\\
&=&i\la0|T\bar{\pi}_i(x)\bar{\pi}_i^\dagger(0)|0\ra\scs
\en
where the ${\bar \pi}_i$ denotes a free field. The $i\epsilon$-contribution
is generated by a damping factor 
$-\epsilon\int d^4x \sum_i\pi_i^\dagger(x)\pi_i(x)$ in the action. To ease
notation, we always omit this term in the following.
As is seen from the integral representation (\ref{propagator_0}), the 
propagator
$G^0_{NR,i}$ vanishes for negative times, from where we conclude that
the free fields $\bar{\pi}_i$ annihilate the vacuum. As a result of
this, the Lagrangian $\lagnr$ conserves the number
of pions. This fact is, of course, built in  - a term like
 e.g. $(\pi_0^\dagger)^4\pi_0^2 +h.c.$ would violate this rule.

We now discuss Green functions in the presence of interactions, and
start the discussion for the case where  the Coulomb term is absent, 
$\alpha=0$.
Again, the relevant Green functions may be evaluated in the standard
manner through the path integral. First, we note that all tadpole
diagrams vanish in (split) dimensional regularization, and we adhere
in the following to this convention. The only corrections to
the two-point function are mass insertions, generated by ${\cal L}_D$.
Summing these up, we obtain 

\eq\label{full propagator}
G_{NR,i}(x)=(2\pi)^{-4}\int\frac{d^4p \, \, e^{-ipx}}{\omega_i({\bf
p})-p^0}\, ; \quad\quad \omega_i({\bf p})=\sqrt{M_{\pi^i}^2+{\bf p}^2}\, ,
\en
with
\eq
(i\partial_t-\sqrt{M_{\pi^i}^2-\triangle})G_{NR,i}(x)=-\delta^4(x)\per
\en

Next we consider the four-point functions, relevant for
elastic $\pi\pi$ scattering. To be specific, we consider the process
\eq
\pi^+(p_1)\pi^-(p_2)\rightarrow \pi^+(p_3)\pi^-(p_4)\per
\en
The corresponding connected Green function is
\eq
G_{NR}^{\pm;\pm}(p_3,p_4;p_1,p_2)&=&
i^4\int d^4x_1\ldots d^4x_4 \,\, 
e^{-i(p_1x_1+p_2x_2-p_3x_3-p_4x_4)}\nonumber\\
&\times&
\la0|T\pi_+(x_3)\pi_-(x_4)\pi_+^\dagger(x_1)\pi_-^\dagger(x_2)|0\ra_c\per
\en

Some of the diagrams generated by the
interactions are displayed in figure \ref{fig:strongloops}. There 
are two classes of
diagrams: Mass insertions generated by ${\cal L}_D$, and bubbles
generated by ${\cal L}_S$ . The perturbative calculation is simply
performed by an expansion in the number of loops and 
mass insertions. The reason why this expansion is meaningful 
 is the following. 
In the CM frame  $P^\mu=p_1^\mu+p_2^\mu=(P^0,{\bf 0})$,
the elementary 
``building blocks'' to calculate a diagram with any
number of bubbles are given by the loop integral
\eq\label{building-blocks}
J_{i}(P^0)&=&\int\frac{d^Dl}{(2\pi)^Di}\,\,
\frac{1}{M_{\pi^i}+{\bf l}^2/(2M_{\pi^i})-P^0+l^0}\,\,
\frac{1}{M_{\pi^i}+{\bf l}^2/(2M_{\pi^i})-l^0}
\nonumber\\[2mm]
&=&\frac{i M_{\pi^i}}{4\pi}\,\,(M_{\pi^i}(P^0-2M_{\pi^i}))^{1/2}\quad\quad
{\rm at}~~D\rightarrow 4 \,\,\,, P^0>2M_{\pi^i}\per
\en
The function $J_{i}$ is analytic in the complex $P^0$ plane,
cut along the real axis for $P^0>2M_{\pi^i}$. As shown below,
the contribution to the scattering matrix element is obtained 
 by putting $ P^0=2w_+({\bf p})$,
 where ${\bf p}$ denotes the pion three momentum in the CM frame.
The loop integral is then purely imaginary. In the case where charged
pions are running in the loop, the integral is of order 
 $|{\bf p}|^{1/2}$ near threshold. For neutral pions in the loop,
it is proportional to $(\mpp-\mpz)^{1/2}$ at the threshold
$P^0=2\mpp$. In case that some of the vertices contain derivatives
 (denoted by the full circle in figure \ref{fig:strongloops}c), and/or
when mass insertions occur in internal lines, additional factors 
 $|{\bf p}|$ and/or  $(\mpp-\mpz)$ appear. As a result of this, 
the expansion in the number of loops and  mass insertions is at the
same time an expansion in $|{\bf p}|$ and in the isospin breaking 
parameter $\mpp-\mpz$.
We conclude that, to calculate the scattering amplitude at a given
order in the momenta or in the isospin breaking parameters, only a
finite number of diagrams  need to be  considered.

We now to discuss  mass insertions on the external lines.
These have to be summed up in order to generate the correct pole
positions at $p_i^0=\sqrt{\mpps+{\bf p}_i^2}$. On the other hand, the
insertions in the internal lines can be treated perturbatively. For a
detailed discussion of this issue we refer the interested reader to 
Ref.~\cite{Lamb}. The Green function is then of the form
\eq\label{reduction1}
G_{NR}^{\pm;\pm}(p_3,p_4;p_1,p_2)=\prod_i(\omega_+({\bf
p}_i)-p_i^0)^{-1}
R^{\pm;\pm}(p_3,p_4;p_1,p_2)\per
\en
The scattering amplitude is obtained from $R^{\pm;\pm}$ by putting all
momenta on their mass shell,
\eq\label{reduction2}
R^{\pm;\pm}|_{p_i^0=w_+({\bf p}_i)} &=&\la \pi^+({\bf
p}_3)\,\pi^-({\bf p}_4)
\,\mbox{out}|\pi^+({\bf p}_1)\,\pi^-({\bf p}_2) \,\mbox{in} \ra_c \nonumber\\
&\doteq&i(2\pi)^4\delta^4(p_1+p_2-p_3-p_4)T_{NR}^{\pm;\pm}({\bf
p}_3,{\bf p}_4;{\bf p}_1,{\bf p}_2)\scs
\en
with normalization
\eq
\la \pi^+({\bf p}_1)|\pi^+({\bf p}_2)\ra = (2\pi)^3\delta^3({\bf p}_1-{\bf
p}_2)\per
\en 
Note that, since the two-point function has residue equal to one, the wave
function renormalization constants are unity as well.

A formula similar to (\ref{reduction1}), (\ref{reduction2}) holds for any 
 $2\rightarrow 2$ scattering process
\eq
\pi^i(p_1)\pi^k(p_2)\rightarrow \pi^l(p_3)\pi^m(p_4)\per
\en
The corresponding relativistic amplitudes are related to the non 
relativistic ones through
\eq\label{R-NR}
T^{lm;ik}_R({\bf p}_3,{\bf p}_4; {\bf p}_1,{\bf p}_2)
=4\left\{w_i({\bf p}_1)w_k({\bf p}_2)w_l({\bf p}_3) w_m({\bf
p}_4)\right\}^{1/2}
T^{lm;ik}_{NR}({\bf p}_3,{\bf p}_4; {\bf p}_1,{\bf p}_2 )\per
\en

In the following, we denote the total and relative momenta by
\eq
{\bf P}={\bf p}_1 + {\bf p}_2, 
\hspace*{1cm} {\bf p}=\frac{1}{2}({\bf p}_1-{\bf p}_2)\per
\en
Unless stated otherwise, we consider  scattering processes always in the CM
frame ${\bf P}$=0.

\subsection{The low-energy constants  - matching}
We  discuss the role of the low-energy constants
$c_i$ that occur in the effective theory. We first
consider the equal mass case $M_{\pi^+}=M_{\pi^0}=M_\pi$, discard the
Coulomb interaction ${\cal L}_C$, and 
 write the
corresponding LECs as $\bar{c}_1$, $\bar{c}_2$, $\bar{c}_3$ and $\bar{c}_4$. 
The matrix elements for the scattering processes
\eq
&&\pi^+\pi^-\rightarrow \pi^+\pi^-\, ,\nonumber\\
&&\pi^+\pi^-\rightarrow \pi^0\pi^0\scs \pi^0\pi^0\rightarrow
\pi^+\pi^-\, ,\nonumber\\
&&\pi^0\pi^0\rightarrow \pi^0\pi^0\, ,
\en
 are obtained from the residue of
the relevant four-point function, as we just discussed.
 Each contribution consists of a product of  loop functions $J_i$,
 including vertices with derivatives and/or mass
insertions. Near threshold, the
 loop expansion
generates  a power series in ${\bf p}$,
\eq\label{expansion1}
\bar{T}_{NR}=\bar f_0+|{\bf p}|\bar f_1+{\bf p}^2\bar f_2+O(|{\bf p}|^3)\, ,
\en
where $\bar{T}_{NR}$ denotes a generic elastic scattering amplitude.
The coefficients $\bar f_i$ depend on the constants $\bar{c}_i$, on the pion
mass $M_\pi$ and on the scattering angle.
 The threshold amplitude $\bar f_0$  receives a contribution from 
the  tree graph
alone. By use of the relation (\ref{R-NR}), we therefore find that
\eq\label{match}
4M_{\pi}^2 \bar{c}_1&=&T^{\pm;\pm}_R\, , \nonumber\\
 8M_{\pi}^2 \bar{c}_2&=& T^{00;\pm}_R =
T^{\pm;00}_R\, , \nonumber\\
16 M_{\pi}^2 \bar{c}_3&=&T^{00;00}_R\scs 
\en
 where $\bar{T}^{\pm;\pm}_R $ stands for the
 relativistic matrix element, 
 evaluated at threshold in the equal mass
case, in the absence of electromagnetic interactions.

We have not yet
specified what relativistic theory we are considering - the relations 
(\ref{match}) are true for any of these. Let us consider  QCD,
 and  represent the  threshold amplitudes
through the relevant scattering lengths in the isospin symmetry limit
$m_u=m_d$. We then have
\eq\label{matchc_i}
3M_{\pi}^2 \bar{c}_1&=&4\pi(2 a_0+a_2)\scs\nonumber\\
3M_{\pi}^2 \bar{c}_2&=&4\pi(a_2 - a_0)\scs\nonumber\\
3M_{\pi}^2 \bar{c}_3&=&2\pi(a_0+2 a_2)\scs
\en
where $M_{\pi}$ denotes the pion mass in QCD at $m_u=m_d$. 
These relations are  true to
all orders  in the chiral expansion. 

\subsection{Matching with the chiral expansion}
There is a second possibility to perform the matching. Namely, one may
arrange the couplings $c_i$ such that ${\cal L}_{NR}$ reproduces the
chiral expansion of the relativistic amplitude to a given order in 
chiral perturbation theory. To arrive at the relevant expression, it
is sufficient to work out the chiral expansion of the threshold
amplitudes at a given order in the chiral expansion and to compare the 
result with (\ref{match}). At order $p^2$, the chiral 
amplitudes are
\eq\label{tree-rel}
T(\pi^+\pi^-\rightarrow\pi^+\pi^-)&=&\frac{s+t-2M^2}{F^2}\, ,
\nonumber\\[2mm]
T(\pi^+\pi^-\rightarrow\pi^0\pi^0)&=&-\, \frac{s-M^2}{F^2}\, ,
\nonumber\\[2mm]
T(\pi^0\pi^0\rightarrow\pi^0\pi^0)&=&\frac{s+t+u-3M^2}{F^2}\, ,
\en
where
\eq
M^2=(m_u+m_d)B\, \, \, , \, B=\frac{1}{F^2}|\langle 0|\bar{u}u|0\rangle|\, ,
\en
and where $F$ is the pion decay constant in the chiral limit $m_u=m_d=0$.
In the isospin symmetry limit $\alpha=0, m_u=m_d$, the parameter
$M$ is further related to the  pion mass through
\eq
M_\pi^2=M^2+O(p^4)\, .
\en
The symbols $s,~t,~u$ denote the standard Mandelstam variables. It
follows that
\eq
{\bar c}_1=\frac{1}{2F^2}+\cdots\scs\quad\quad
{\bar c}_2=-\frac{3}{8F^2}+\cdots\scs\quad\quad
{\bar c}_3=\frac{1}{16F^2}+\cdots \scs
\en
where the ellipses denote higher-order terms in the quark mass
expansion. With these values of the LECs, the tree graphs of ${\cal
L}_{NR}$ reproduce the leading order in the chiral expansion of the
threshold amplitudes. Similarly, ${\bar c}_4$ can be related to the momentum
dependence of $T(\pi^+\pi^-\rightarrow\pi^0\pi^0)$,
\eq
{\bar c}_4=\frac{1}{32 F^2 M_{\pi}^2}+\cdots\per
\en

\subsection{Including the Coulomb interaction}
\label{sec:including_Coulomb}
We now consider Green functions at order $\alpha$, and relax the
 equal mass condition for the pions. There are two classes of diagrams: The
 first one contains the same diagrams as $\bar{T}_{NR}$, but now evaluated 
 at $M_{\pi^+}\neq M_{\pi^0}$, and with  couplings $c_i$ that  depend
on $\alpha$ and $m_u-m_d$, see below. The second class contains 
diagrams  with  one virtual Coulomb photon. Feynman graphs
 where the Coulomb photon is attached in such a manner that pions must
 propagate in time in order to connect the two vertices - the self-energy graph is
 an example - all vanish. This is because  one may close the contour
of integration over the zero-component of the photon momentum  in a
 half-plane where there is no singularity in the propagators.
 Since the self-energy diagrams vanish, 
the mass parameters 
$M_{\pi^+}$ and  $M_{\pi^0}$ in the Lagrangian 
may be identified with the physical masses. The two-point functions for the
 charged and neutral pion field 
are therefore still given by the expression (\ref{full propagator}).
We now consider  virtual Coulomb-diagrams that are  built
 from diagrams displayed in Fig. \ref{fig:Coulomb}. The 
crosses  in the figure denote mass
 insertions. We  evaluate the contributions from
 Figs. \ref{fig:Coulomb}b,c  and start the discussion with the Coulomb 
vertex diagram
 Fig.~\ref{fig:Coulomb}b, with no mass insertion.
 After
integration over the zero component of the loop momentum, the
  integral to be evaluated is
\eq\label{Vc_I}
V_c({\bf p},P^0)=e^2\int\frac{d^d{\bf l}}{(2\pi)^d}\frac{1}{|{\bf l}|^2}
\frac{1}{P^0-2\mpp -({\bf p}-{\bf l})^2/\mpp}\, , \, \quad d=D-1 . 
\en
The contribution to the scattering amplitudes is obtained by
evaluating this expression at $P^0=2w_+({\bf p})$. The result is
\eq\label{Vc}
V_c({\bf p}, 2w_+({\bf p}))=-\frac{\pi\alpha\mpp}{4 |{\bf p}|} 
- i\alpha\theta_c +O(|{\bf p}|,d-3)\scs
\en
where
\eq\label{tc}
\theta_c=\frac{\mpp}{2|{\bf
p}|}\mu^{d-3}\left\{\frac{1}{d-3}-\frac{1}{2}\bigl[\ln{4\pi}+\Gamma'(1)\bigr]
+\ln{\frac{2|{\bf p}|}{ \mu}}\right\}
\en
is the infrared-divergent Coulomb phase \cite{piP-Atom}.

Next, we consider  the two-loop diagram Fig.~\ref{fig:Coulomb}c,
 omitting mass insertions.
Again integrating over the zero components of
the loop momenta, the corresponding amplitude is expressed in terms of 
\eq\label{Bc_I}
B_c(P^0)=\frac{e^2}{(2\pi)^{2d}}\int\frac{d^d{\bf l}_1}{P^0-2\mpp -{{\bf
l}_1}^2/\mpp}\, 
\frac{1}{|{\bf l}_1-{\bf l}_2|^2}\frac{d^d{\bf l}_2}
{P^0-2\mpp-{\bf l}_2^2/\mpp}\per
\en
Evaluating this expression at $P^0=2w_+({\bf p})$, we find
\eq\label{def_Lambda}
B_c(2w_+({\bf p}))&=&-\frac{\alpha \mpps}{8\pi}
\left\{\Lambda(\mu) +2\ln{\frac{2|{\bf p}|}{\mu}}-1-i\pi\right\} 
+O(|{\bf p}|,d-3)\, ,\nonumber\\
\Lambda(\mu)&=&\mu^{2(d-3)}\left\{\frac{1}{d-3}-\ln{4\pi}
-\Gamma'(1)\right\}\per
\en
The ultraviolet divergences in diagrams that contain $B_c$ are
removed in the standard manner by adding counterterms to the
Lagrangian ${\cal L}_{NR}$. For the consistency of the method it is
important to notice that the diagrams obtained 
by adding mass insertions and/or using vertices with derivative
couplings, are  suppressed by powers of  momenta with respect
to the leading terms $B_c$ and $V_c$. They will not be 
needed in the following.

The structure of the elastic scattering amplitudes up to and including terms
of order $\alpha$ is now as follows. First we note that, since the
propagators are not affected by the self-energy diagrams, the reduction
 formulae (\ref{reduction1}), (\ref{reduction2}) are still valid. We write the
 generic  scattering amplitude as
\eq
T_{NR}=T_{NR}^0+\alpha T^1_{NR} +O(\alpha^2)\, ,
\en
where $T^1_{NR}$ contains one virtual Coulomb photon.
The expansion of the first term in powers of the center
of mass momenta is as in Eq. (\ref{expansion1}), with  coefficients 
 $f_i$ that  now also depend on the pion mass difference, 
and on $\alpha$ through the
 coupling constants $c_i$. Omitting the tree contribution
 from one-Coulomb exchange displayed in Fig. \ref{fig:Coulomb}a,
 we write the second term as
\eq
T_{NR}^1=\frac{M_{\pi^+}}{|{\bf p}|}\left\{ g_0+|{\bf p}|g_1 
+|{\bf p}|\ln\,\frac{2|{\bf p}|}{M_{\pi^+}}\,g_2+{\bf p}^2 g_3
+\cdots\right\}\, .
\en
The coefficients $g_i$ contain in general infrared  divergences,
generated by the vertex diagram $V_c$.
Otherwise, the structure of the $g_i$
is again the same as the one of the coefficients $f_i$ in $T^0_{NR}$.
Power counting also works in this general case: there is only a finite number
of diagrams that contribute to a given coefficient $f_i$ or $g_i$.
Finally,  the relation to the amplitude in the underlying
relativistic theory is again given by Eq. (\ref{R-NR})\footnote{Both the
 relativistic and non-relativistic amplitudes must be evaluated 
by using the same infrared regulator, such that  the Coulomb-phase 
can be identified on both sides. We find it convenient to use
 dimensional regularization.}.

One may  perform the matching to the chiral expansion also in this
general case.  First, we note that 
 $\alpha$ is then counted as a quantity of order $p^2$. 
Second, the chiral
representation (\ref{tree-rel}) is valid at order $p^2$ also in the
presence of electromagnetic interactions, provided that one 
 i) identifies the quantity $M$ in
(\ref{tree-rel}) with the neutral pion mass, and ii) 
 adds the
one-photon exchange amplitude in  the
$\pi^+\pi^-\rightarrow\pi^+\pi^-$   channel. Let us match the amplitudes 
at order $p^2$. Counting powers of $F^2$, it is easy to see that loop diagrams
 $T^1_{NR}$ do not contribute - the 
matching relations become \cite{Bern1}
\eq\label{c_i}
&&c_1=\frac{1}{2F^2}(1+\kappa)+\cdots,\quad\quad\hspace*{.2cm}
c_2=-\frac{3}{8F^2}\biggl(1+\frac{\kappa}{6}\biggr)+\cdots,
\nonumber\\[2mm]
&&c_3=\frac{1}{16F^2}+\cdots,\quad\quad\quad\quad\quad  
c_4=\frac{1}{32F^2M_{\pi^0}^2}(1-2\kappa)+\cdots,
\en
with $\kappa=M_{\pi^+}^2/M_{\pi^0}^2-1$. The ellipses stand either for
terms at $o(\kappa)$, or higher-order contributions  in the chiral expansion.
The terms of order $\kappa$ are proportional to $\alpha$ at this
order in the chiral expansion - Eq.~(\ref{c_i})
displays the $\alpha$-dependence of the couplings $c_i$ mentioned above.
 
This concludes our discussion of the evaluation of Green functions in
the non-relativistic theory.

\setcounter{equation}{0}
\section{Pionium in the non-relativistic framework}
\label{sec:bound-states}

The bound states and their decays are most conveniently described in a
Hamiltonian framework. The effective theory discussed above renders
the pertinent calculations rather straightforward, as we will now show.

\subsection{Hamiltonian and Fock space}
The non-relativistic Lagrangian ${\cal L}_{NR}$ gives rise to the following 
Hamiltonian,
\eq\label{Ham_full}
{\bf H}&=&{\bf H}_0+{\bf H}_D+{\bf H}_C+{\bf H}_S=
{\bf H}_0+{\bf H}_C+{\bf V}, \nonumber\\[2mm]
{\bf H}_\Gamma&=&\int d^3{\bf x} \, {\cal H}_\Gamma(0,{\bf x})\, ;  \hspace*{.3cm} 
\Gamma=0,D,C,S,\nonumber\\[2mm]
{\cal H}_0&=& \sum_{i=\pm, 0}\,\pi_i^\dagger\biggl( 
M_{\pi^i}-\frac{\triangle}{2M_{\pi^i}}\biggr)\pi_i,\quad\quad
{\cal H}_D=-\sum_{i=\pm, 0}\,\pi_i^\dagger
\biggl(\frac{\triangle^2}{8M_{\pi^i}^3}+\cdots\biggr)\pi_i,
\nonumber\\[2mm]
{\cal H}_C&=&4\pi\alpha(\pi_-^\dagger\pi_-)\triangle^{-1}
(\pi_+^\dagger\pi_+)\, ,
\nonumber\\[2mm]
{\cal H}_S&=&-c_1\pi_+^\dagger\pi_-^\dagger\pi_+\pi_-
-c_2[\pi_+^\dagger\pi_-^\dagger(\pi_0)^2+{\rm h.c.}]
-c_3\,(\pi_0^\dagger\pi_0)^2
\nonumber\\[2mm]
&-&c_4[\pi_+^\dagger\stackrel{\leftrightarrow}
{\triangle}\pi_-^\dagger(\pi_0)^2+\pi_+^\dagger\pi_-^\dagger\pi_0
\stackrel{\leftrightarrow}{\triangle}\pi_0+{\rm h.c.}]\,.
\en
It is convenient to introduce creation and annihilation operators,
\eq\label{commutation}
[{\bf a}_i({\bf p}),{\bf a}_k^\dagger({\bf q})] &=&
(2\pi)^3\delta^{3}({\bf p}-{\bf q})\delta_{ik}\, \, ; \, \,
i,k=\pm,0\scs\nonumber\\
\pi_i(0,{\bf x})&=&\int d\nu({\bf p})e^{i{\bf p}{\bf x}}
{\bf a}_i({\bf p})\, ;\quad\quad
d\nu({\bf p})= \frac{d^3{\bf p}}{(2\pi)^3}\per
\en
The free Hamiltonian becomes
\eq
{\bf H}_0 =  \int d\nu({\bf p})\sum_i\biggl(M_{\pi^i}+\frac{{\bf p}^2}{2M_{\pi^i}}
\biggr)a_i^\dagger({\bf p}) a_i({\bf p})\scs
\en
and the propagator, evaluated with the free fields
\eq
\bar{\pi}_i(x^0,{\bf x})=
e^{i{\bf H}_0 x^0}\bar \pi_i(0,{\bf x})e^{-i{\bf H}_0x^0}\scs
\en
of course agrees with the expression Eq. (\ref{propagator_0}).
 We will also need the two-particle states with zero total charge,
\eq\label{twoparticlestates}
|{\bf P},{\bf p}\rangle_+={\bf a}_+^\dagger({\bf p}_1)\,
{\bf a}_-^\dagger({\bf p}_2)\, |0\rangle\scs
|{\bf P},{\bf p}\rangle_0={\bf a}_0^\dagger({\bf p}_1)\,
{\bf a}_0^\dagger({\bf p}_2)\, |0\rangle\per
\en
In terms of these, the unperturbed pionium ground state is given by
\eq\label{Coulomb-state}
|\Psi_0,{\bf P}\rangle=\int d\nu({\bf q})\,\Psi_0({\bf q})\,
 |{\bf P},{\bf q}\rangle_+\, ,
\en
where  $\Psi_0({\bf q})$ is the
 Coulomb wave function in the momentum space
\eq\label{Coulomb-wf}
\Psi_0({\bf q})=\frac{(64\pi \gamma^5)^{1/2}}{({\bf q}^2+\gamma^2)^2}\, ,
\quad\quad
\gamma=\frac{1}{2}\,\alpha M_{\pi^+}\scs
\en
and
\eq\label{Coulomb-eq}
&&({\bf H}_0+{\bf H}_C)\,|\Psi_0,{\bf P}\rangle=
(E_0+E_{CM})|\Psi_0,{\bf P}\rangle\scs\nonumber\\
&&E_0=2M_{\pi^+}\biggl(1-\frac{\alpha^2}{8}\biggr), \hspace*{.5cm} 
E_{CM}=\frac{{\bf P}^2}{4M_{\pi^+}}\per
\en
The perturbation ${\bf V}$ renders the ground state unstable. We
discuss in the remaining part of this article how the corresponding
width can be evaluated.

\subsection{Resolvents - the master equation}
\label{sec:resolvents}
To determine the width of the ground state, we have considered in
Ref. \cite{Bern1} 
the scattering amplitude in the neutral
channel,
$\pi^0\pi^0\rightarrow\pi^0\pi^0$, and determined the position of its
poles in the complex energy plane. Here, we instead make use of 
resolvents. While the two descriptions are perfectly equivalent,
 we find that the use of the resolvent renders the calculations even
 simpler. We begin the discussion with the quantity 
\eq
{\bf G}_C(z)=\frac{1}{z-{\bf H}_0-{\bf H}_C}\scs
\en
 whose  matrix elements between the charged states 
(\ref{twoparticlestates}) develop poles at the position of the energy
levels of the unperturbed pionium. To remove the CM momentum of the
matrix elements, we introduce the notation
\eq
({\bf q}|{\bf r}(z)|{\bf p})=
\int d\nu({\bf P}) \, 
_+\langle{\bf P},{\bf q}|{\bf R}(z)|{\bf 0},{\bf p}\rangle_+\scs
\en
where ${\bf R}(z)$ denotes any operator in Fock space.
 One can now easily relate the matrix element of  ${\bf G}_C$ to 
Schwinger's Green function~\cite{Schwinger},
\eq\label{Schwinger}
({\bf q}| {\bf g}_C(z)|{\bf p})&=& 
\frac{(2\pi)^3\delta^{3}({\bf q}-{\bf p})}
  {E-{\bf q}^2/M_{\pi^+}}-\frac{1}{E-{\bf q}^2/M_{\pi^+}}\,
  \frac{4\pi\alpha}{|{\bf q}-{\bf p}|^2}\,\frac{1}{E-{\bf p}^2/M_{\pi^+}}
  \nonumber\\[2mm]
  &-& \frac{1}{E-{\bf q}^2/M_{\pi^+}}\,\,
  4\pi\alpha\eta I(E;{\bf q},{\bf p})\,\,\frac{1}{E-{\bf p}^2/M_{\pi^+}}\, ,
\en
with
\eq\label{Schwinger-I}
I(E;{\bf q},{\bf p})=
\int_0^1\frac{x^{-\eta} dx}
  {[({\bf q}-{\bf p})^2x+\eta^2/\alpha^2(1-x)^2(E-{\bf q}^2/M_{\pi^+})
    (E-{\bf p}^2/M_{\pi^+})]}\, ,
\en
where $\eta=\frac{1}{2}\,\alpha\,(-E/M_{\pi^+})^{-1/2}$ and
$E=z-2M_{\pi^+}$.
This function has poles at $\eta=1,2,\ldots \per$. In order to
calculate the position of the poles in the real world, with ${\bf
V}\neq 0$, we consider the full resolvent 
\eq
{\bf{G}}(z)=\frac{1}{z-{\bf H}}\per
\en
 Expanding  in powers of the perturbation
${\bf V}$, one finds that ${\bf G}$  satisfies the equation
\eq
{\bf G}&=&{\bf G}_C+{\bf G}_C \mathbold{ \tau} {\bf G}_C\scs\nonumber\\
\mathbold{\tau}&=& {\bf V}+{\bf V}{\bf G}_C\mathbold{ \tau}\per
\en
We remove the ground state
singularity from ${\bf G}_C$,
\eq
\bar{{\bf G}}_C={\bf G}_C\left\{{\bf 1}-
\int 
 d\nu({\bf P})|\Psi_0,{\bf P}\ra\la\Psi_0,{\bf P}|\right\} \scs
\en
introduce 
\eq\label{tau-def}
\bar{\mathbold{\tau}}={\bf V}+{\bf
V}\bar{G}_C\bar{\mathbold{\tau}}\scs
\en
and find for ${\bf G}$ the representation
\eq
{\bf G}=\bar{\bf G}_C+\bar{\bf G}_C\bar{\mathbold{\tau}}\bar{\bf G}_C
+(1+\bar{\bf G}_C\bar{\mathbold{\tau}}){\bf \Pi}_0
(1+\bar{\mathbold{\tau}}\bar{\bf G}_C)\scs
\en
where
\eq\label{eqpoles}
{\bf \Pi}_0=\int\frac{ d\nu({\bf P})|\Psi_0,{\bf
P}\ra\la\Psi_0,{\bf P}|}
{z-E_{CM}-E_0-
(\Psi_0|\bar{\mathbold{\tau}}(z;{\bf P})|\Psi_0)}\scs
\en
and
\eq\label{eqtau}
(\Psi_0|\bar{\mathbold{\tau}}(z;{\bf P})|\Psi_0)=
\int d\nu({\bf P'})\la\Psi_0,{\bf P'}|
\bar{\mathbold{\tau}}(z)|\Psi_0,{\bf P}\ra
\per
\en
The singularity
generated by the ground state pole is absent in the barred
quantities. Therefore, the
pertinent pole must occur through a zero in the denominator of the
expression
(\ref{eqpoles}). In the CM frame, the relevant
 eigenvalue equation to be solved is
\eq\label{key}
z -E_0-
(\Psi_0|\bar{\mathbold{\tau}}(z)|\Psi_0) =0\scs
\en
where the  matrix element denotes the quantity on the left-hand
side of Eq. (\ref{eqtau}), evaluated in the CM frame ${\bf P}=0$.

The  master equation (\ref{key})  is a compact form of the
conventional Rayleigh-Schr\"{o}dinger perturbation theory. Note  that
 it  fixes the convergence domain of the perturbation theory:
the theory is applicable  as long as  the energy-level shift 
 does not become comparable to the distance between the ground-state and
the first radial-excited Coulomb poles. Equation (\ref{key}) is valid for
 a general potential- containing e.g. the interaction with the
transverse photons - since in the derivation, we did not use the explicit
form of the interaction Hamiltonian in (\ref{Ham_full}). 

\subsection{Singularity structure of the resolvent}
We find it instructive to shortly discuss the analytic structure of
the matrix elements of the resolvent ${\bf G}$, and the location of the
shifted ground state pole.
 First, from (\ref{key}), it is seen that
this pole will occur at the same position for any channel.
Second, it is expected on general grounds that the pole 
will move to the second Riemann sheet.
Indeed,  consider the
operator $\bar{\mathbold{\tau}}$ in the second iterative approximation
\eq
\bar{\mathbold{\tau}}={\bf V}+{\bf V}\bar{{\bf G}}_C {\bf V} +O({\bf
V}^3)\per
\en
To evaluate the matrix element between charged states as required, we
insert a complete set of neutral states in the second term. The
eigenvalue equation becomes
\eq\label{second}
z=E_0+\frac{M_{\pi^+}^3\alpha^3}{8\pi}\left\{-c_1-2c_2^2J_0(z)+\cdots\right\}\scs
\en
where $J_0$ denotes the loop integral (\ref{building-blocks}). This function
has a branch point at $z=2\mpz$, and its imaginary part has the same sign as
the imaginary part of $z$ throughout the cut $z$-plane. Therefore, the
 equation (\ref{second}) has  no solution on the first
Riemann sheet.
 On the other hand, if we analytically continue 
 $J_0$ from the upper rim of the cut to the second Riemann sheet,
we find that  a zero at
\eq\label{example}
z&=&{\rm Re}z+i\,{\rm Im} z\scs\nonumber\\
{\rm Re}z&=&E_0- \frac{\alpha^3 \mpp^3 }{8\pi}\, c_1+\cdots\scs\quad\quad
{\rm Im}z=-\frac{\alpha^3 \mpp^3 \mpz}{16\pi^2}\,\rho^{1/2}c_2^2+\cdots\, ,
\en
 with  
 $\rho=2M_{\pi^0}(M_{\pi^+}-M_{\pi^0}-M_{\pi^+}\alpha^2/8)$.

The imaginary part is of order $\delta^{7/2}$. We 
demonstrate below that it is the only term at this order. Using
Eq. (\ref{matchc_i}) and $\Gamma=-2{\rm Im}z$, one recovers
\cite{Labelle} the leading order result (\ref{eqleading}).

Similar arguments   apply to all the other pole positions.
 [Of course, in order to correctly describe the new 
positions of the exited energy levels, our original Lagrangian ${\cal L}_{NR}$
 must  be enlarged.] We
conclude that the 2-particle matrix elements of ${\bf G}$ are 
analytic functions in the
complex $z$-plane, cut along the real axis for ${\rm Re} z > 2\mpz$. The
 poles are located 
on the second Riemann sheet. 

\setcounter{equation}{0}
\section{Pionium decays}
\label{sec:pionium}

\subsection{Perturbative solution of the bound-state equation}
In order to find the solution to Eq. (\ref{key}) at order 
$\delta^{9/2}$, it is convenient 
to reduce the equation (\ref{tau-def})
to a one-channel problem with an effective potential ${\bf W}$. We 
 use a projector $\varrho$ on the two-particle states $|{\bf P},{\bf q}\ra_+$,

\eq\label{projectors}
{\varrho}&=&\int d\nu({\bf P})d\nu({\bf q})|{\bf P},{\bf q}{\ra_+}
 {_+\la}{\bf P},{\bf q}|\scs\nonumber\\
\varrho_0&=&{\bf 1}-\varrho\, ,
\en
and find in the standard manner
\eq\label{eq-w++}
\varrho {\bar{\mathbold{\tau}}}\varrho &=&
\varrho {\bf W}\varrho\,  + \varrho 
{\bf W}\varrho\, 
{\bar{\bf G}_C}\,\varrho {\bar{\mathbold{\tau}}}\varrho 
\scs\nonumber\\
{\bf W}&=&{\bf V}+{\bf V}\varrho_0{\bar{\bf G}_C}
\left\{{\bf 1}-\varrho_0{\bf V}
\varrho_0{\bar{\bf G}_C}\right\}^{-1}\varrho_0{\bf V}\per
\en 
This is result is still perfectly general. In the case considered
here,  one may simplify the expression for the
effective potential, replacing $\bar {\bf G}_C$ by ${\bf G}_0
=(z-{\bf H}_0)^{-1}$,
\eq\label{w++_0}
{\bf W}={\bf V}+{\bf V}\varrho_0{\bf G}_0\left\{
{\bf 1}-\varrho_0{\bf V}\varrho_0{\bf G}_0\right\}^{-1}\varrho_0{\bf V}\, .
\en 

The matrix elements of the effective potential can be expanded in powers of
momenta, because there are no nearby singularities. Specifically, we write
\eq\label{w++}
({\bf q}|{\bf w}(z)|{\bf p})
&=&(2\pi)^3\delta^3({\bf q}-{\bf p})\,
\biggl(-\frac{{\bf p}^4}{4M_{\pi^+}^3}+\cdots\biggr)
\nonumber\\[2mm]
&+&{ w}(z)+{ w}_1(z){\bf p}^2+{ w}_2(z){\bf q}^2+
{ w}_3(z){\bf pq}+\cdots\, .
\en
If we now iterate the equation~(\ref{eq-w++}), at the order of accuracy we
are working, the decay width of the $\pi^+\pi^-$ atom 
$\Gamma_{2\pi^0}=-2{\rm Im}\, z$ is given by
\eq\label{Gamma}
\Gamma_{2\pi^0}=-\frac{\alpha^3M_{\pi^+}^3}{4\pi}\, {\rm Im}\, { w}\,
(1+2\, {\rm Re}\,{ w}\,\langle\bar{\bf g}_C(E_0)\rangle\,) +O(\delta^5)\, ,
\en
where ${ w}={ w}(E_0)$, and
\eq\label{int-g}
\langle\bar {\bf g}_C(E_0)\rangle=\int\frac{d^d{\bf q}}{(2\pi)^d}\,
\frac{d^d{\bf p}}{(2\pi)^d}\, ({\bf q}|\bar {\bf g}_C(E_0)|{\bf p})\, .
\en
In order to calculate this integral, one needs to define
 Schwinger's Green function  in $d$ dimensions.
In  field theory, the Fourier transform of the Coulomb potential 
in $d$ dimensions is given by
exactly the same expression as in 3 dimensions - consequently, the first two
terms in the representation~(\ref{Schwinger}) are also valid at 
$d\neq 3$.  For the last term, the integral is convergent, 
and we may work at  $d=3$.
The integral is then equal to
\eq\label{int-g-ans}
\langle\bar {\bf g}_C(E_0)\rangle&=&
\frac{\alpha M_{\pi^+}^2}{8\pi}\,\xi\, ,\quad\quad
\xi=2\ln\alpha-3+\Lambda(\mu)+\ln\frac{M_{\pi^+}^2}{\mu^2}\, .
\en
The quantities ${\rm Re}\,{ w}$ and ${\rm Im}\,{ w}$ 
can be determined from  iterations of Eq.~(\ref{eq-w++}) to the needed 
accuracy,
\eq\label{Rew-Imw}
{\rm Re}\, { w}&=&-c_1\, ,
\nonumber\\[2mm]
{\rm Im}\, { w}&=&-\frac{M_{\pi^0}}{2\pi}\,\rho^{1/2}\,
\biggl(1+\frac{5\rho}{8M_{\pi^0}^2}\biggr)\,
\bigl( c_2-2\rho c_4\bigr)^2
\biggl(1-\rho\,\frac{M_{\pi^0}^2c_3^2}{4\pi^2}\biggr)\, . 
\en 
We have now expressed the decay width of the $\pi^+\pi^-$ atom in terms of 
the non-relativistic couplings $c_1\cdots c_4$. 
It remains to determine the relevant combination of these couplings 
from the matching of the relativistic and non-relativistic
amplitudes.

\subsection{Matching to the threshold amplitude}
\label{sec:matching}

We determine the non-relativistic couplings that enter the
expression for the decay width through Eq. (\ref{Rew-Imw}), and start the
discussion with the couplings $c_1$ and $c_3$.  
These  contribute to the
decay width at order $\delta^{9/2}$, because  $\rho$ counts as a quantity of
order $\delta$. Therefore, these two couplings are needed at order
$\delta^0$ (no isospin breaking), as a result of which we  may replace them by
the isospin symmetric quantities $\bar{c}_1$ and $\bar{c}_3$ in 
Eqs. (\ref{matchc_i}). It remains  
to determine the combination of the couplings
$c_2$ and $c_4$ that enter (\ref{Rew-Imw}). As we will now show,
 it suffices for this purpose to calculate the real part of 
the $\pi^+\pi^-\rightarrow\pi^0\pi^0$
amplitude at order $\delta$ in the non-relativistic and in the relativistic
theories.

Whereas the Hamiltonian framework is very convenient to discuss the energy
spectrum, it is more convenient to calculate scattering amplitudes in the
Lagrangian framework discussed in section \ref{sec:eff-theory}. 
In the relativistic theory, the on-shell amplitude for  
$\pi^+\pi^-\rightarrow\pi^0\pi^0$ contains infrared
singularities that exponentiate~\cite{Yennie},
\eq\label{B-factor}
&&T^{00;\pm}_{R}=\exp(\alpha B^{00;\pm})\,\hat T^{00;\pm}_{R}, 
\hspace*{.5cm} 
B^{00;\pm}=-2\pi\,\int\frac{d^Dl}{(2\pi)^Di}\, 
\frac{{\cal P}^2(l)}{l^2}, \nonumber\\[2mm]
&&{\cal P}(l) = \frac{2p_1+l}{l^2+2lp_1}+\frac{2p_2-l}{l^2-2lp_2}\, ,
\en
where 
$p_1,~p_2$ denote the 4-momenta of incoming $\pi^+$ and $\pi^-$ mesons.
In Ref.~\cite{Yennie} it is demonstrated that - using a photon mass as
an infrared regulator -  the residual amplitude 
$\hat T^{00;\pm}_{R}$ is free of infrared singularities. 
Here we assume that the same is true in dimensional regularization.
We find
\eq\label{B-thr}
B^{00;\pm}=i\theta_c+\frac{\pi M_{\pi^+}}{4|{\bf p}|}
-\frac{3}{2\pi}+O(|{\bf p}|,D-4)\, ,
\en
where $\theta_c$ is defined by Eq.~(\ref{tc}). The infrared
divergences cancel in the real part of $B^{00;\pm}$ at threshold, whereas the 
imaginary part is divergent at $D\rightarrow 4$. 

One may verify that at order $\alpha$, exactly the same
divergent Coulomb phase appears in the non-relativistic amplitudes.
Indeed, if one performs the calculation at  $d\neq 3$ and splits off the
phase according to
\eq\label{tD}
T_{NR}^{00;\pm}({\bf q},{\bf p})&=&e^{i\alpha \theta_c}
{\hat T}_{NR}^{00;\pm}({\bf q},{\bf p})\, ,
\en
then there are no infrared singularities in the amplitude 
${\hat T}_{NR}^{00;\pm}({\bf q},{\bf p})$
at threshold in the limit  $d\rightarrow 3$, at order $\delta$. For
the real part, we find\footnote{
Note that in Ref.~\cite{Bern1}, the non-relativistic scattering amplitude is
defined with an opposite sign.}
\eq\label{bar-tD}
{\rm Re}\, {\hat T}_{NR}^{00;\pm}({\bf q},{\bf p})=
\frac{B_1}{|{\bf p}|}\, +B_2\,\ln\frac{2|{\bf p}|}{M_{\pi^+}}
+\frac{1}{4M_{\pi^+}^2}\, {\rm Re}\, A^{+-00}_{\rm thr}+o({\bf p})\, ,
\en
where
\eq\label{B1-B2}
B_1=\frac{\pi\alpha M_{\pi^+}}{2}\, c_2+o(\delta)\, ,\quad\quad
B_2=-\frac{\alpha M_{\pi^+}^2}{2\pi}\, c_1c_2+o(\delta)\, .
\en
The singular contributions $\sim 1/|{\bf p}|, \ln{|{\bf p}}|$ 
are generated by  the exchange of one Coulomb photon (see
Figs.~\ref{fig:Coulomb}b,\ref{fig:Coulomb}c).
At $O(\delta)$, the constant term in Eq.~(\ref{bar-tD}) is equal to
\eq\label{ReA}
\frac{1}{4M_{\pi^+}^2}\, {\rm Re}\, A^{+-00}_{\rm thr}=
2c_2-4M_{\pi^0}^2\kappa\biggl( c_4+\frac{c_2c_3^2M_{\pi^0}^2}{8\pi^2}\biggr)
+\frac{\alpha M_{\pi^+}^2}{4\pi}\,\biggl( 1-\Lambda(\mu) 
-\ln\frac{M_{\pi^+}^2}{\mu^2}\biggr)\, c_1c_2 ,
\en 
where $\Lambda(\mu)$ is given by Eq.~(\ref{def_Lambda}), and $\kappa$ is 
the same as in Eq.~(\ref{c_i}). The ultraviolet divergence contained in 
 $\Lambda(\mu)$ may be 
absorbed in the renormalization of the coupling $c_2$. This procedure at the 
same time eliminates the ultraviolet divergence in the expression for
the  decay width.

In the following, we assume that - up to and including terms of
order  $\delta$ - the relativistic amplitude does
have the same singularity structure - as a function of the momentum
${\bf p}$ - 
as the non-relativistic amplitude (\ref{bar-tD}). We can then match the
non-relativistic expression to the relativistic one in the standard
manner, using Eq. (\ref{R-NR}). The quantity  
${\rm Re}\, A^{+-00}_{\rm thr}$ in (\ref{bar-tD}) corresponds 
to the one introduced
in Ref.~\cite{Knecht}, in the context of the relativistic theory
(modulo the Coulomb phase, which does, however, not contribute to the
amplitude at order $e^2p^2$.).
 The logarithmic singularity is absent in the amplitude at order 
$e^2p^2$ at which the calculations in Ref.~\cite{Knecht} were carried out - 
it first emerges at order $e^2p^4$~\cite{Roig}, see Appendix 
\ref{app:threshold}.
Finally, the relation~(\ref{ReA})  represents
the matching condition  between the regular part 
of the relativistic  $\pi^+\pi^-\rightarrow\pi^0\pi^0$ scattering
 amplitude at threshold and the pertinent combination of  
non-relativistic coupling constants $c_i$.

\subsection{General expression for the decay width}

Substituting the results of the matching into the expression for the
$\pi^+\pi^-$ atom decay width~(\ref{Gamma}), and
using~(\ref{int-g-ans}) and~(\ref{Rew-Imw}),  we obtain
\eq\label{Gamma-fin}
&&\Gamma_{2\pi^0}=\frac{2}{9}\,\alpha^3 p^\star{\cal A}^2(1+K)\, ,\quad\quad
{\cal A}=-\frac{3}{32\pi}\,{\rm Re}\, A^{+-00}_{\rm thr}+o(\delta)\, ,
\nonumber\\[2mm]
&&K=\frac{\kappa}{9}\, 
(a_0+2a_2)^2-\frac{2\alpha}{3}({\rm ln}\alpha-1)(2a_0+a_2)+o(\delta)\, 
,\nonumber\\
&&p^\star=(\mpps-\mpzs-\frac{1}{4}\mpps\alpha^2)^{1/2}\, .
\en
This is the general expression for the $\pi^+\pi^-$ atom decay width, valid at
 next-to-leading order in isospin breaking, and to all orders
in the chiral expansion. Note that all mention of the non-relativistic
theory has disappeared in the final result that relates the observable
quantity (the decay width) to the relativistic scattering amplitude at
threshold.

The primary objective of the DIRAC experiment is to measure the difference
$a_0-a_2$ of the $S$-wave $\pi\pi$ scattering lengths that are defined in the
isospin-symmetric world. The expression~(\ref{Gamma-fin}) is not yet suited
 for this purpose, because it relates the width to the
scattering amplitude at threshold. This quantity contains 
the combination $a_0-a_2$ we are looking for, together with isospin breaking 
contributions.  One has to evaluate these  and subtract them
 from the measured amplitude.
ChPT allows one to achieve this goal - order by order in the expansion in the
quark mass.

\subsection{Amplitude at $O(e^2p^2)$}

The normalization of the quantity ${\cal A}$ is chosen such that, in
the isospin symmetry limit, it coincides with the
difference  $a_0-a_2$ of the $S$-wave scattering lengths. In the general
case, we expand the amplitude in powers of the isospin breaking
parameters $\alpha$ and $m_u-m_d$,
\eq\label{chiral-A}
{\cal A}=a_0-a_2+h_1(m_d-m_u)^2+h_2\alpha+o(\delta)\, .
\en
This decomposition is true irrespective of the chiral expansion. The
scattering lengths as well as the coefficients $h_i$ are functions of
the quark mass $\hat{m}$ and of the renormalization group invariant
scale of QCD. 
What is the meaning of $a_0-a_2$ in the presence of isospin-violating
interactions?
To clarify the issue, we consider
the expression ${\cal A}$ at leading order in the chiral
expansion. From (\ref{tree-rel}), we find 
\eq\label{A-leading}
{\cal A}=\frac{3}{32\pi F^2}(4 M^2_{\pi^+}-M^2) 
+ O(p^4,e^2p^2)\, .
\en
To bring this into the form (\ref{chiral-A}),
 we note that, in the isospin-symmetry limit $m_u=m_d,\alpha=0$,
the scattering lengths can be expanded in powers of
the pion mass, defined to be the position of the pole 
in the  correlator  of two axial currents. It is an algebraically
 perfectly legitimate procedure to identify this mass with the
charged pion mass. We adhere in the following to this procedure, in order
to agree with the standard conventions in ChPT. The expression for the
 difference of the scattering lengths then reads
\eq
a_0-a_2=\frac{9M^2_{\pi^+}}{32\pi F^2} +O(p^4)\, .
\en
Comparing this with (\ref{A-leading}),
 we find
\eq
{\cal A} = a_0-a_2 +\frac{3 (M^2_{\pi^+}-M^2)}{32 \pi F^2}
+O(p^4,e^2p^2)\, .
\en
 From this result, we may read off the coefficient  $h_2$ at
 leading order in the chiral expansion,
\eq
h_2= \frac{3 (M^2_{\pi^+}-M^2)}{32 \alpha\pi F^2}
+O(\hat{m})\, .
\en
[To be precise, the first term on the right-hand side of this equation
should be evaluated at $\alpha = 0$. To ease notation, we omit this
request here and in the following]. On the other hand, the above
calculation is not accurate enough to determine $h_1$  at leading
order, because for this purpose, the amplitude is needed at order $p^4$.
This procedure may obviously be carried out
  order by order in the chiral
expansion  - all that is needed is the 
chiral expansion of the scattering amplitude at threshold, 
at $m_u\neq m_d$, $\alpha\neq 0$.
 As a result of this, the quantities $h_i$ are represented  as a
 power series in the quark mass $\hat{m}$ (up to logarithms). 

The evaluation of  the amplitude for $\pi^+\pi^-\rightarrow\pi^0\pi^0$
has been  carried out  at $O(p^4,e^2p^2)$ 
in Ref.~\cite{Knecht}. This result allows us therefore to
determine the coefficient $h_1$ ($h_2)$  at order $p^0$ ($p^2$).
 Some remarks are in order.

\begin{itemize}
\item[i)]{In Ref.~\cite{Knecht}, the scattering amplitude has been
    evaluated at $m_u=m_d$. For our purposes, the expression 
    for generic $m_u$ and $m_d$ is needed.
    On the other hand,
    up to and including terms of order $p^4$, the strong
     amplitude does not contain $m_u-m_d$ terms. The only source
    for such contributions is the tree graph,
     where $M^2$  is expressed in terms
    of the  neutral pion mass. The generalization of the
    result Ref.~\cite{Knecht} to the unequal
    mass case is therefore  straightforward.}
\item[ii)]{The normalization point in Ref.~\cite{Knecht} is chosen to be the
    neutral pion mass. According to our definitions, we have to normalize all
    low-energy constants at the charged pion mass. The terms that
    emerge from the shift of the normalization point are proportional to
    $\Delta_\pi=M_{\pi^+}^2-M_{\pi^0}^2$
     and are included in the expressions given below.} 
\end{itemize}

The rest is then straightforward. We find
\eq\label{h_i}
&&h_1=O(\hat m)\, ,\nonumber\\
&& h_2=\frac{3\Delta_\pi^{em}}{32\pi\alpha F^2}\,\biggl( 1+
\frac{M_{\pi^+}^2}{12\pi^2F^2}\, \biggl[\frac{23}{8}+\bar l_1
+\frac{3}{4}\,\bar l_3\biggr]\biggr) +\frac{3M_{\pi^+}^2}{256\pi^2F^2}
\, p(k_i)\,  + O(\hat{m}^2)\, ,
\en
where $p(k_i)$ stands for the following combination of the 
electromagnetic low-energy constants~\cite{Knecht},
\eq\label{pk_i}
p(k_i)=-30+9\bar k_1+6\bar k_3+2\bar k_6+\bar k_8+\frac{4}{3}\, Z
(\bar k_1+2\bar k_2+6\bar k_4+12\bar k_6-6\bar k_8)\, ,
\en
and
\eq\label{Delta_em} 
\Delta_\pi^{em}=\Delta_\pi\biggr|_{m_u=m_d}\, ,\quad\quad
Z=\frac{\Delta_\pi^{em}}{8\pi\alpha F^2}\, .
\en
The quantities $\bar k_i$ denote again the running coupling constants
$k_i^r(\mu)$ at scale $\mu=M_{\pi^+}$. Note that according to our counting,
the quantity $L_\pi=\ln(M_{\pi^+}^2/M_{\pi^0}^2)$ introduced in
Ref.~\cite{Knecht} is of order $\delta$ and hence does not contribute to
$h_2$. Further,  $F$ may be  expressed through $F_\pi$ 
according to~\cite{ChPTlit1}
\eq\label{F_pi}
F_\pi=F\,\biggl(1+\frac{M_{\pi^+}^2}{16\pi^2F^2}\,\bar l_4+O(\hat m^2)\biggr)
\, .
\en
 For the numerical
analysis, one has to specify the values of the low-energy constants that 
enter the expression for $h_2$. We are not aware 
of an estimate for the  $SU(2)\times SU(2)$  couplings $\bar k_i$.
On the other hand, the corresponding couplings $K_i$ in 
$SU(3)\times SU(3)$~\cite{Urech}  have been estimated by 
invoking e.g. sum rules or  a resonance saturation hypothesis 
\cite{Baur,Moussallam,Bijnens}.
In order to use this information, we need to relate the couplings
$\bar{k}_i$ to their $SU(3)$ counterparts $K_i$. 
In Appendix \ref{app:mapping}, we show that 
\eq\label{kK}
p(k_i)=P(K_i)-8Z\bar l_4\, ,
\en
where
\eq\label{PK_i}
P(K_i)&=&\frac{128\pi^2}{3}\,\biggl[-6(K_1^r+K_3^r)+3K_4^r-5K_5^r+K_6^r
+6(K_8^r+K_{10}^r+K_{11}^r)\biggr]
\nonumber\\[2mm]
&-&(18+28Z)\ln\frac{M_{\pi^+}^2}{\mu^2}
-2Z\biggl(\ln\frac{m_sB_0}{\mu^2}+1\biggr)-30\, ,
\en
and where $K_i^r$ denote the running couplings introduced in \cite{Urech}.
Taking into account this relation, we may rewrite the formula for the
width in the following form,
\eq\label{G-eps}
\Gamma_{2\pi^0}=\frac{2}{9}\alpha^3p^\star\,(a_0-a_2+\epsilon)^2(1+K)\, ,
\en
with
\eq\label{eps}
\epsilon&=&\frac{3\Delta_\pi^{em}}{32\pi F_\pi^2}\,
\biggl(1+\frac{M_{\pi^+}^2}{12\pi^2F_\pi^2}\,
\biggl[\frac{23}{8}+\bar l_1+\frac{3}{4}\, \bar l_3\biggr]\biggr)
+\frac{3\alpha M_{\pi^+}^2}{256\pi^2F_\pi^2} P(K_i)\nonumber\\[2mm]
&&+ O(\hat{m}(m_u-m_d)^2, \alpha \hat{m}^2) + o(\delta)\,.
\en
The quantity $K$ is given in Eq.~(\ref{Gamma-fin}).
 
\subsection{Numerical analysis}

In the numerical evaluation of the lifetime, we use for $a_0$ 
and $a_2$ the values from the recent analysis in
Ref.~\cite{Colangelo},
$a_0=0.220\pm 0.005$, $a_2=-0.0444\pm 0.001$, $a_0-a_2=0.265\pm
0.004$. To evaluate the correction $\epsilon$,   
 we first recall that the non-electromagnetic 
part of the pion mass difference is tiny, of  
order $\sim 0.1~{\rm MeV}$~\cite{Reports}. 
Therefore, we identify $\Delta_\pi^{em}$  with the experimentally measured 
total shift $\Delta_\pi$. 
Further, in the calculations we replace $m_s B_0$ by
$M_{K^+}^2-M_{\pi^+}^2/2$, according to our definition of the 
isospin symmetry
limit. The values used for the low-energy constants in the strong
sector are
$\bar l_1=-0.4\pm 0.6$, $\bar l_3=2.9\pm 2.4$ \cite{Colangelo}. For $K_i^r(\mu)$, we use the
values given by Baur and Urech in  Ref.~\cite[Table 1]{Baur}: 
$K_1^r=-6.4,~
 K_3^r=6.4,~
 K_4^r=-6.2,~
 K_5^r=19.9,~
 K_6^r=8.6,~
 K_8^r=K_{10}^r=0,~
 K_{11}^r=0.6$ (in units of $10^{-3}$). 
We evaluate $P(K_i)$ at scale $\mu=M_\rho$.
Further, we attribute an uncertainty $2/16\pi^2$ - that stems from 
dimensional arguments - to each  $K_i^r$. The values of $K_i^r$
obtained both by Moussallam~\cite{Moussallam} and by Bijnens and 
Prades~\cite{Bijnens}, lie then within the uncertainties attributed. 
The same is true, if the saturation is assumed not at scale $\mu=M_\rho$,
but somewhere within the interval $0.5~{\rm GeV}\leq\mu\leq 1.0~{\rm GeV}$.
Finally, we use $F_\pi=92.4~{\rm MeV}$.
Adding the uncertainties in $l_1,l_3, a_0, a_2$ and in $K_i$
quadratically, we obtain
\eq\label{eps-K}
\epsilon=(0.61\pm 0.16)\times 10^{-2}\, ,\quad\quad
K=(1.15\pm 0.03)\times 10^{-2}\, ,
\en
or
\eq\label{correction}
\Gamma_{2\pi^0}=\frac{2}{9}\, \alpha^3 p^\star\, (a_0-a_2)^2(1+\delta_\Gamma)
\, ,\quad\quad \delta_\Gamma=(5.8\pm 1.2)\times 10^{-2} \, .
\en
This amounts to a six percent correction to leading-order formula by
Deser {\it et al}~\cite{Deser}. 
In the total decay width $\Gamma$, the decay into $2\pi^0$ is by far
the dominating mode. For example, the decay width into a $2\gamma$ pair, 
which is the first subleading mode in $\delta$ counting, 
is $\Gamma_{2\gamma}={\alpha^5}\,
M_{\pi^+}/4$~\cite{DIRAC,Hammer} at leading order in the
$\delta$-expansion, as a result of which one has
$\Gamma_{2\gamma}/\Gamma_{2\pi^0} \simeq 3\times 10^{-3}$. For this reason,
one may safely identify $\tau_{2\pi^0}$ with the total lifetime,
\eq\label{tau}
\tau\doteq \Gamma_{2\pi^0}^{-1}=(2.9\pm 0.1)\times10^{-15}~{\mbox s}\, .
\en 
We add the following remarks concerning these numbers.
\begin{itemize}
\item
The bulk part in the uncertainty in the lifetime is due to the 
uncertainty in the difference of the
scattering lengths $a_0-a_2$, which  results in $\pm 0.085\times 10^{-15}$
s.
\item
The uncertainties in the constants $K_i$ increase this to 
$\pm 0.091\times 10^{-15}$ s. Including the remaining uncertainties does not 
change this number in the digits displayed.
\item
 {The numbers in Eqs.~(\ref{eps-K})-(\ref{tau}) differ from the
corresponding ones in our
previous paper~\cite{Bern2}, because the present  values of the
scattering lengths, of $\bar{l}_1$ and  of $F_\pi$  differ from the
 ones used there.
 The above values of $a_0,a_2$  are the result of a 
 complete analysis~\cite{Colangelo} at order $p^6$ -
 they replace the ones  used in \cite{Bern2}, taken from the preliminary 
 numerical result cited in  Ref.~\cite{pipi-NUCL}. 
 The present value of $\bar{l}_1$ is based 
 the same  analysis~\cite{Colangelo}.
 The bulk part in 
 the change of the lifetime  is of course due to
the updated  value of $a_0-a_2$, because this combination of
scattering lengths enters the expression for the decay rate at 
leading order.}
\item
The vacuum polarization correction to the lifetime,
 that is not taken into account here, amounts~\cite{Labelle} to a
 contribution of  $-0.01\times 10^{-15}$ s. 
\end{itemize}

We expect that the higher-order
contributions to the $\pi^+\pi^-$ atom decay width in ChPT are 
 negligibly small.
Consequently, an accurate determination of $a_0-a_2$ from a precise lifetime
measurement is indeed feasible.

\setcounter{equation}{0}
\section{Summary}
\label{sec:conclusions}

\begin{itemize}
\item[i)]{We have considered  decays of the $\pi^+\pi^-$
atom  in its ground state. Aside from a kinematical factor, the
decay rate can be expanded in powers of the isospin breaking
parameters $\alpha$ and $(m_u-m_d)^2$. It is convenient to
book these parameters  as terms of order $\delta$.}
\item[ii)]{To calculate the leading and next-to-leading order
term in the $\delta$-expansion of the width, we have constructed a 
non-relativistic
Lagrangian that describes the low-energy interactions of pions and photons.
 In this framework, the matrix elements of the
resolvent $1/(z-{\bf H})$ develop poles on the second Riemann sheet in
the complex $z$-plane. The positions of the poles are related to the
energy levels and widths in the standard manner.
By using Feshbach's technique, we have derived
the master equation~(\ref{key}) for the position of the ground-state pole.}
\item[iii)] {On the basis of this equation, we have calculated the
decay width of the
ground state of pionium in terms of the parameters of the
non-relativistic Lagrangian. At leading and next-to-leading order
 in the $\delta$-expansion, 
only the channel $A_{\pi^+\pi^-}\rightarrow \pi^0\pi^0$ is open.
Furthermore, at this order of accuracy,  transverse photons do not
contribute - the relevant Lagrangian becomes then very simple, see
Eq. (\ref{Lagr_full}). Matching the non-relativistic amplitude to 
the relativistic one, we
have then expressed the decay width in terms of the relativistic scattering
amplitude, up to terms that vanish faster than $\delta^{9/2}$. The
relevant formula is displayed in Eq. (\ref{Gamma-fin}).}
\item[iv)] {At this stage, one may  invoke ChPT,  which allows one to 
expand the isospin breaking terms in powers of the quark mass, and thus to get
contact with measurable quantities. The result is given in
Eq. (\ref{G-eps}), that displays the width in terms of the combination
$a_0-a_2$ of $S$-wave scattering
lengths, and a correction that we have calculated at order
$\alpha$ and $\hat{m}\alpha$. The quark mass difference shows up only at
order $(m_u-m_d)^2\hat{m}$. We expect this term to be completely
negligible. The recently determined values\cite{Colangelo} of 
 the $\pi\pi $ $S$-wave scattering lengths gives
\eq
\tau=(2.9\pm 0.1)\times10^{-15}s.
\en
}
\item[v)]
{Since the isospin breaking  corrections at order
$\alpha$ and $\hat{m}\alpha$ are small, we expect that chiral corrections
at higher order as well as higher-order terms in isospin breaking 
are irrelevant for  data on the lifetime obtained in the foreseeable future.
}

\end{itemize}

\section*{Acknowledgments}
 We are grateful to 
H. Leutwyler, L.L. Nemenov and J. Schacher
for useful discussions. V.E.L. thanks the University of Bern for hospitality.
This work was supported in part by the Swiss National Science
Foundation, and by TMR, BBW-Contract No. 97.0131  and  EC-Contract
No. ERBFMRX-CT980169 (EURODA$\Phi$NE).

\appendix

\renewcommand{\thesection}{\Alph{section}}
\renewcommand{\theequation}{\Alph{section}\arabic{equation}}
\setcounter{section}{0}
\section{General non-relativistic Lagrangian}
\label{app:Lagrangian}
\setcounter{equation}{0}

In this Appendix, we outline  general rules for the construction
of a non-relativistic
Lagrangian that describes low-energy interactions of pions and
photons. The Lagrangian does not contain terms that correspond to 
transitions between sectors with different number of heavy
particles (pions), since these belong to hard processes
and are hidden in the couplings of the
non-relativistic Lagrangian. For this reason, in order to describe 
$\pi\pi$ scattering in the non-relativistic framework, it suffices to
consider  Lagrangians in the sectors with one or two pions
(including any number of photons).  
The theory must be  invariant 
under space rotations, $C$, $P$, $T$ and gauge transformations. 
On the other hand,  due e.g. to the presence of photons, the Lagrangian is not
invariant under Galilei transformations.  
The appropriate building blocks 
are provided by the covariant derivatives of the charged
pion fields
\eq\label{covariant}
D_t\, \pi_\pm=\partial_t\,\pi_\pm\mp ieA_0\,\pi_\pm\, ,\quad\quad
{\bf D}\, \pi_\pm= \nabla\,\pi_\pm\pm ie{\bf A}\,\pi_\pm\, ,
\en
 and the electric and magnetic fields
\eq\label{EB}
{\bf E}=-\nabla A_0-\dot{\bf A}\, ,\quad\quad {\bf B}={\rm rot}\, {\bf A}\, .
\en
For the neutral pion field, the covariant derivative coincides with the
ordinary one. 

The Lagrangian consists of an infinite tower of operators with  increasing
mass dimension. All possible operators allowed by the symmetries must be
included. In particular, in the  one-pion sector,
 the Lagrangian is given by
\eq\label{Lagr-1}
{\cal L}_1&=&\frac{1}{2}\, ({\bf E}^2-{\bf B}^2)+
\pi_0^\dagger\,\biggl\{i\partial_t-M_{\pi^0}+\frac{\triangle}{2M_{\pi^0}}
+\frac{\triangle^2}{8M_{\pi^0}^3}+\cdots\biggr\}\,\pi_0
\nonumber\\[2mm]
&+&\sum_\pm\pi_\pm^\dagger\,\biggl\{iD_t-M_{\pi^+}
+\frac{{\bf D}^2}{2M_{\pi^+}}+\frac{{\bf D}^4}{8M_{\pi^+}^3}
+\cdots\biggr\}\,\pi_\pm\, ,
\en
complemented, e.g., with all possible non-minimal couplings 
containing ${\bf E}$ and
${\bf B}$, that we have  not explicitly displayed.
At  tree level, this
Lagrangian reproduces the relativistic result for the scattering
amplitude $\pi+m\gamma\rightarrow\pi+n\gamma$
at $O(e^{m+n})$, to all orders in the momentum expansion.

In the  two-pion sector of  zero total charge, one 
has to construct the
operators that contain four pion fields and any number of photon fields.
The lowest-order Lagrangians with zero and two space derivatives are given by
\eq\label{L0}
{\cal L}_2^{(0)}&=&c_1\pi_+^\dagger\pi_-^\dagger\pi_+\pi_-
+c_2(\pi_+^\dagger\pi_-^\dagger\pi_0\pi_0+{\rm h.c.})
+c_3\pi_0^\dagger\pi_0^\dagger\pi_0\pi_0\, ,
\\[2mm]\label{L2}
{\cal L}_2^{(2)}&=&c_4\biggl\{(\pi_+^\dagger\stackrel{\leftrightarrow}
{\bf D}^2\pi_-^\dagger)(\pi_0\pi_0)+(\pi_+^\dagger\pi_-^\dagger)(\pi_0
\stackrel{\leftrightarrow}{\bf D}^2\pi_0)+{\rm h.c.}\biggr\}
\nonumber\\[2mm]
&+&c_5\biggl\{(\pi_+^\dagger\stackrel{\leftrightarrow}
{\bf D}^2\pi_-^\dagger)(\pi_+\pi_-)+{\rm h.c.}\biggr\}
+c_6\biggl\{[\pi_+^\dagger\pi_+]\stackrel{\leftrightarrow}{\bf D}^2
[\pi_-^\dagger\pi_-]\biggr\}
\nonumber\\[2mm]
&+&c_7\biggl\{(\pi_0^\dagger\stackrel{\leftrightarrow}
{\bf D}^2\pi_0^\dagger)(\pi_0\pi_0)+{\rm h.c.}\biggr\}\, ,
\en
where $u\stackrel{\leftrightarrow}{\bf D}^2\!\! v\doteq 
u\, {\bf D}^2v+v\, {\bf D}^2u$.
Note that the couplings $c_i$ are not necessarily real, as a result of
which the
Lagrangian is not, in general,  hermitian (see below).
Again, we do not display explicitly  non-minimal couplings
 that, apart from covariant space derivatives, contain
the vectors ${\bf E}$ and ${\bf B}$. Moreover, we do not display
 covariant time derivatives, or higher-order 
Lagrangians ${\cal L}_2^{(4)},~{\cal L}_2^{(6)},~\cdots$
which contain $4,~6,~\cdots$ space derivatives. In the 
absence of photons, the Lagrangian ${\cal L}_2^{(0)} +{\cal L}_2^{(2)}$ is
 equivalent to the one given in 
Ref.~\cite{Soto2}. 

In the isospin symmetry limit $\alpha=0,~m_u=m_d$, the following relations 
hold for the corresponding couplings $\bar c_1,\cdots \bar c_7$,
\eq\label{isospinlimit}
\bar c_3=\frac{1}{2}\,(\bar c_1+\bar c_2)\, ,\quad\quad
\bar c_7=\frac{1}{2}\,(\bar c_4+\bar c_5+\bar c_6)\, .
\en
In the sectors with one or two pions, the Lagrangian is
therefore given by
\eq\label{full-lag}
{\cal L}={\cal L}_1+{\cal L}_2^{(0)}+{\cal L}_2^{(2)}+\cdots\, ,
\en
where the ellipses stand for  non-minimal, or  higher-dimensional
operators or higher-order terms with covariant time derivatives. On 
the mass shell, the latter terms are eliminated by using
the equation of motion (EOM). However, they need to be
included if one decides to  renormalize Green functions~\cite{Lamb}.
At tree level, the Lagrangian~(\ref{full-lag})
reproduces the relativistic result for the scattering amplitude 
$k\pi+m\gamma\rightarrow k\pi+n\gamma,~k=1,2$
at $O(e^{m+n})$, to all orders in the momentum expansion. 

The scattering amplitudes in the non-relativistic theory are related -
through the reduction formula - to the residues of the pertinent
Green functions in a standard manner.
[Note that, in the non-relativistic theory, one 
has  to sum up all insertions $\sim {\bf p}^4,~{\bf p}^6,\cdots$ in the
external legs of the pions, in order to ensure that the poles 
sit at 
the correct position, according  to the relativistic dispersion law 
$p_i^0=(M_{\pi^i}^2+{\bf p}_i^2)^{1/2}$. On the other hand, insertions
 in the internal
lines are treated perturbatively. Fore more 
details, we refer the reader to Ref.~\cite{Lamb}.]
The loop corrections to the Green functions
are then calculated in a standard manner, by using  Feynman 
diagrammatic techniques, with one important modification.
It is well known that in the non-relativistic theory, in the 
presence of light particles (photons), the Feynman
integrals should be properly butchered, in order to avoid  contributions
from the loop momenta at a hard scale - otherwise, loop corrections to
the Green functions would lead to a breakdown of the counting rules in the
non-relativistic theory. A suitable procedure built on top
of the Feynman rules in the non-relativistic theory is provided
by the so-called threshold expansion~\cite{Beneke} (see also~\cite{Manohar}),
that enables one to
disentangle the contributions coming from different regions of loop momenta,
by expanding the integrands - in the dimensional regularization scheme
- in all
possible small kinematical variables. Here, we  adopt a simple and physically
transparent formulation of such a procedure~\cite{Soto-private}.
First, we put a momentum cutoff on all 3-dimensional Feynman integrals,
after having performed the integrals over all zero components of
virtual four-vectors (by eventually using split dimensional
regularization). 
Then we  choose the cutoff mass to be much smaller than the hard scale,
given by the pion mass. Next, we expand the integrands both in external and
integration momenta. In the presence
of the cutoff, this is a well-defined procedure. At the last step, we
remove the
cutoff and calculate the integrals, with the expanded integrands, in 
dimensional regularization. This sequence of steps systematically removes 
the hard-momentum contribution from the integrals, which at low energies 
is given by a polynomial in the external momenta.

All couplings in the
non-relativistic Lagrangian are determined by matching to the
relativistic theory. In the presence of photons, 
the non-relativistic Lagrangian is, in general,  not 
hermitian - because these constants are not real. This is due to  the fact
that in the non-relativistic approach, one has  shielded some of the
intermediate states - those with  masses below the two-pion threshold -
that appear in the relativistic theory and that
belong to the class of hard processes in the non-relativistic terminology.
The imaginary part of such diagrams then contributes to the imaginary part of
the couplings of the non-relativistic Lagrangian. For example, the decay of
the $\pi^+\pi^-$ atom into two photons in the relativistic theory is 
described - at leading order in $\alpha$ and in $\hat{m}$ - 
by the imaginary parts of the diagrams depicted in Fig.~\ref{fig:2-gamma}.
These diagrams
are not present in the non-relativistic theory. On the other hand,
they contribute to the
imaginary part of the coefficient $c_1$ at order
$O(e^4)$. 

The next remark concerns  power counting. In fact, we have three
different types
of power counting in our theory:

\begin{itemize}
\item[i)] Non-relativistic power counting. The Green functions -
calculated at a
  fixed order in an expansion in the coupling  $e$ - are expanded
in powers of external 3-momenta of the particles,
 and in the mass difference $M_{\pi^+}-M_{\pi^0}$ in the manner
  described in Sections~\ref{sec:alpha=0} and~\ref{sec:including_Coulomb}.
\item[ii)] Chiral power counting. After matching to ChPT,
the couplings are given in a form of a series in the quark masses and $e$.
The coefficients contain  the  low-energy constants (LECs) of ChPT. 
This procedure is
systematic in the sense that matching at higher chiral order does not
affect the result obtained at lower orders. 
\item[iii)] Counting the isospin-breaking parameter $\delta$. 
After matching to ChPT,
the couplings in the non-relativistic Lagrangian can be rewritten as
\eq\label{couplings}
c_i=c_i^{(0)}+\alpha c_i^{(1)}+(m_d-m_u)^2c_i^{(2)}
+o(\delta)\, ,\quad\quad
c_i^{(n)}=c_i^{(n)}(M_{\pi^+}^2,{\rm LECs})\, ,
\en
where, by convention, we have defined the isospin-symmetric world with
$\alpha=0$, and $m_d=m_u$ as the one
in which the common mass of the pion triplet coincides with the mass of the
charged pion in the real world
[Note that in the relativistic pion scattering amplitudes,  odd powers of
$m_d-m_u$ never appear.].
Consequently, the powers of $\delta$ in the 
expansion of the Green functions around the isospin symmetry limit stem
from different sources. The explicit powers are due to the coupling to 
photons, and the implicit powers are encoded in the couplings of the
Lagrangian, as well as the charged and neutral pion mass difference.
In the calculations, one has to carefully keep track of all these sources 
of corrections in a given order in $\delta$.
\end{itemize}

The Lagrangian (\ref{full-lag}) contains an infinite number of operators.
In actual calculations, only a few of them are needed. In particular,
we will make it plausible that,  in order to calculate the pionium
decay width at $O(\delta^{9/2})$, it suffices to work with the
Lagrangian given in Eq.~(\ref{Lagr_full}). The arguments in favor of
this simplification are provided in the following two Appendixes.

\setcounter{equation}{0}

\section{Threshold expansion and the role of transverse photons}
\label{app:scattering}
\setcounter{equation}{0}
\renewcommand{\thesubsection}{\arabic{subsection}}
We illustrate  the evaluation of the scattering
amplitude in the non-relativistic theory in the presence of
photons. We work in the Coulomb gauge, which allows a
clear-cut separation of Coulomb and transverse photons, 
 and argue that the radiative corrections  to the  
$\pi^+\pi^-\rightarrow\pi^0\pi^0$ scattering amplitude, generated by
transverse photons, vanish at threshold at order $e^2$.

\subsection{Pion self-energy}

We start with the two-point function of the charged pions.
The self-energy correction at $O(e^2)$ due to
the diagram Fig.~\ref{fig:SE}a is given by
\eq\label{two-point}
i\int\, dx\,{\rm e}^{ipx}\,\langle 0|T\pi_\pm(x)\pi_\pm^\dagger(0)|0\rangle=
\frac{1}{M_{\pi^+}+{\bf p}^2/(2M_{\pi^+})-p^0-\Sigma(p^0,{\bf p})}\, ,
\en
\eq\label{self-energy}
\Sigma(p^0,{\bf p})=\frac{e^2}{M_{\pi^+}^2}\int\frac{d^Dl}{(2\pi)^Di}\,\,
\frac{{\bf p}^2-({\bf p}{\bf l})^2/{\bf l}^2}
{-l^2(M_{\pi^+}+({\bf p}-{\bf l})^2/(2M_{\pi^+})
-p^0+l^0)}+O(e^4)\, .
\en
The threshold expansion of the above integral amounts to (we place a hat
above the threshold-expanded quantities)  
\eq\label{hat-SE}
\hat\Sigma(\Omega,{\bf p})&=&
\frac{e^2}{2M_{\pi^+}^2}\int\frac{d^d{\bf l}}{(2\pi)^d}\, 
\biggl({\bf p}^2-\frac{({\bf p}{\bf l})^2}{{\bf l}^2}\biggr)
\frac{1}{|{\bf l}|}
\biggl\{\frac{1}{\Omega+|{\bf l}|}+\biggl(\frac{{\bf p}{\bf l}}{M_{\pi^+}}
-\frac{{\bf l}^2}{2M_{\pi^+}}\biggr)\, \frac{1}{(\Omega+|{\bf l}|)^2}
+\cdots\biggr\}\, ,
\nonumber\\[2mm]
\Omega&=&M_{\pi^+}+\frac{{\bf p}^2}{2M_{\pi^+}}-p^0\, ,
\quad\quad d=D-1\, .
\en
Performing the remaining integration, we obtain
\eq\label{hat-SE-1}
\hat\Sigma(\Omega,{\bf p})&=&\frac{e^2}{2M_{\pi^+}^2}\,{\bf p}^2\,
\Omega^{d-2}\,\frac{\Gamma(d)\Gamma(2-d)}
{(4\pi)^{d/2}\Gamma(1+\frac{d}{2})}+\cdots
\nonumber\\[2mm]
&\rightarrow&
\frac{e^2}{6\pi^2M_{\pi^+}^2}\,{\bf p}^2\,\Omega\,\,
\biggl\{\, L(\mu)+\ln\frac{2\Omega}{\mu}-\frac{1}{3}\biggr\}+\cdots\, ,
\quad\quad{\rm when}~~d\rightarrow 3\, ,
\en
\eq\label{Lmu}
L(\mu)=
\mu^{d-3}\biggl(\frac{1}{d-3}-\frac{1}{2}\,(\Gamma'(1)+\ln 4\pi+1)\biggr)\, .
\en
 As usual, $\mu$ denotes the scale of  dimensional regularization.

In order to remove the divergence from the two-point function, one  introduces
the counterterm 
\eq\label{Delta-L}
\Delta {\cal L}=-\frac{e^2}{6\pi^2M_{\pi^+}^2}\,\sum_{\pm}f_1
\pi_\pm^\dagger {\bf D}^2\biggl(iD_t-M_{\pi^+}
+\frac{{\bf D}^2}{2M_{\pi^+}}\biggr)\pi_\pm\, ,\quad\quad
f_1=L(\mu)+f_1^r(\mu)\, .
\en
The contribution to the two-point function is displayed in Fig.~\ref{fig:SE}b.
The quantity  $f_1^r(\mu)$ denotes the finite,
scale dependent part of the coupling constant $f_1$. 

To calculate the wave function renormalization constant for charged pions, 
one has to reverse the limits~\cite{Lamb}. Namely, we perform the
limit $\Omega\rightarrow 0$
(mass-shell limit in the non-relativistic theory) 
at $d>3$, in order to avoid the infrared singularity. 
Since the ratio  $\hat\Sigma(p^0,{\bf p})/\Omega$ vanishes 
as $\Omega\rightarrow 0$,
 the self-energy diagram Fig.~\ref{fig:SE}a does not
contribute to the wave function renormalization constant. The sole
contribution comes from the counterterm given by Eq.~(\ref{Delta-L}),
\eq\label{Z-charged}
Z_\pm({\bf p}^2)=1-\frac{e^2f_1}{6\pi^2M_{\pi^+}^2}\,\,{\bf p}^2\, .
\en 
Note that $Z_\pm(0)=1$. This feature is due 
to the derivative coupling of transverse photons, and to
 the use of the threshold expansion, which  guarantees that the 
non-relativistic power counting is not altered by loop corrections.

\subsection{Scattering amplitude $\pi^+\pi^-\rightarrow\pi^0\pi^0$ at 
 order $e^2$}
 
We now discuss  the radiative
corrections to the $\pi^+\pi^-\rightarrow\pi^0\pi^0$ scattering amplitude
at order $e^2$, due to transverse photons. We have to consider diagrams with
any number of strong bubbles [including of course the tree diagrams], 
and attach one virtual photon line to these diagrams in all 
possible ways. The photon couples to two pions (the relevant part of 
the Lagrangian is given by Eq.~(\ref{Lagr-1})), 
as well as to the vertices with 
four pions (see Eq.~(\ref{L2})). Further, the photon couples to two pions in a
minimal way, as well as through the non-minimal vertices which contain more
space derivatives acting on the fields.

Some  preliminary remarks are in order. 
After applying the threshold expansion to a given
diagram, one always ends up with a homogeneous integrand, and  naive
power-counting is restored. Since a strong bubble introduces a
suppression factor in a diagram with no photons (see
Section~\ref{sec:eff-theory}), we expect that - even in the presence 
of photons - diagrams containing more strong bubbles will be
 more suppressed, and, 
for a given topology, it suffices to consider diagrams with a minimal  
 number of strong bubbles. The same consideration applies to 
diagrams with non-minimal photon couplings, and to  diagrams with 
derivative couplings in strong four-pion vertices: since power-counting 
holds, we expect that these are suppressed with respect to the diagrams of 
the same topology, but with a minimal number of derivatives in the vertices.

We start with the diagrams where the photon couples only to two pions.
According to the above discussion, we do not consider diagrams with
non-minimal couplings, and restrict ourselves to the non-derivative strong
Lagrangian (\ref{L0}).  
The set of all topologically distinct diagrams with one virtual transverse
photon coupled in a minimal way to two pions, 
is depicted in Fig.~\ref{fig:topologies}. In each class, we single
out a representative with a minimal number of strong loops.

The corrections to the external legs (Fig.~\ref{fig:topologies}a)
vanish at threshold, because $Z_\pm(0)=1$.
Next, we  consider the diagram corresponding to the exchange of a
transverse photon between the initial $\pi^+\pi^-$ pair
(Fig.~\ref{fig:topologies}b). The integral to be
calculated in this case is given by
\eq\label{Sakharov-transverse}
J_{+-\gamma}(|{\bf p}|)&=&-\frac{e^2}{M_{\pi^+}^2}\,\,
\int\frac{d^Dl}{(2\pi)^Di}\,\,\biggl({\bf p}^2
-\frac{({\bf p}{\bf l})^2}{{\bf l}^2}\biggr)\,
\frac{1}{(M_{\pi^+}+({\bf p}-{\bf l})^2/(2M_{\pi^+})-p_+^0+l^0)} 
\nonumber\\[2mm]
&\times&
\frac{1}{l^2(M_{\pi^+}+({\bf p}-{\bf l})^2/(2M_{\pi^+})-p_-^0-l^0)}\, .
\en
One has still to multiply this integral with the purely strong amplitude 
in order to get the contribution of the diagram in 
Fig.~\ref{fig:topologies}b. In this expression,
${\bf p}$ denotes the relative momentum of the $\pi^+\pi^-$ pair in the
CM frame, and $p_+^0,~p_-^0$ are the energies of $\pi^+$ and $\pi^-$
particles. We put the external particles on the mass shell, 
$p_+^0=p_-^0=M_{\pi^+}+{\bf p}^2/(2M_{\pi^+})+O({\bf p}^4)$, and perform the
threshold expansion in the integral. Note that with this procedure, the
integrands also should be expanded in the
$O({\bf p}^4)$ remainder of $p_\pm^0$.
The threshold-expanded integral in Eq.~(\ref{Sakharov-transverse}) 
can be rewritten in the following manner, 
\eq\label{mpole-vertex}
\hat J_{+-\gamma}(|{\bf p}|)&=&\frac{e^2}{M_{\pi^+}}
\int\frac{d^d{\bf l}}{(2\pi)^d}\,
\frac{1}{|{\bf l}|^2}\,\biggl({\bf p}^2
-\frac{({\bf p}{\bf l})^2}{{\bf l}^2}\biggr)\,
\frac{1}{{\bf l}^2-2{\bf p}{\bf l}}+\cdots
\nonumber\\[2mm]
&=&\frac{e^2\,|{\bf p}|}{16M_{\pi^+}}
+\frac{ie^2\, |{\bf p}|}{8\pi M_{\pi^+}}\,\biggl( L(\mu)
+\ln\frac{2|{\bf p}|}{\mu}\biggr)+\cdots\, .
\en
Again, it is seen that this particular contribution vanishes at threshold.

We have also investigated the remaining contributions depicted in
Fig.~\ref{fig:topologies}. All these contributions vanish at threshold.
Moreover, we have considered all topologically distinct sets of diagrams 
where the virtual photons couple to four-pion vertices, depicted
in Fig.~\ref{fig:topologies_4}. Again, in each set we have restricted 
ourselves to the diagram with a minimal   number of strong loops,
and with a minimal number of derivatives in strong and electromagnetic
vertices.  
We have found that the contributions from all these diagrams vanish
at threshold.

To conclude, we have considered all topologically distinct diagrams 
for the scattering process $\pi^+\pi^-\rightarrow\pi^0\pi^0$ in the
non-relativistic theory, where one virtual photon couples in all
possible ways to strong diagrams.
From each class of diagrams, we have singled out the representative
with a minimal number of strong loops, and a minimal number of derivatives in
the vertices. We have checked that each such diagram vanishes at
threshold. Using power-counting, the same is seen to be true for the diagrams
with more loops, and/or higher-order couplings. For this reason,
we expect that all radiative corrections to the
$\pi^+\pi^-\rightarrow\pi^0\pi^0$ scattering amplitude - due to 
transverse photons - vanish at threshold at order $e^2$. 
We therefore neglect the transverse photons in the non-relativistic theory,
while matching the
relativistic and non-relativistic amplitudes at threshold. 

\section{Contribution of transverse photons to the decay width}
\label{app:bound-states}
\setcounter{equation}{0}

We consider the role of transverse photons in the calculation of the
decay width of the $\pi^+\pi^-$ atom. We evaluate their contribution
for several typical diagrams and show that these do not contribute at order
$\delta^{9/2}$. The procedure goes in several steps.

\vspace*{.3cm}

{\bf 1.} As  was mentioned in Section~\ref{sec:resolvents}, 
the master equation~(\ref{key}) for the position of the bound-state
 pole  is valid
 also in the case of a general non-relativistic Lagrangian. Expanding 
in a Taylor
series around $z=E_0$ gives
\eq\label{denom}
z-E_0=\frac{(\Psi_0|\bar{\mathbold{\tau}}(E_0)|\Psi_0)}
{1-\frac{d}{dE_0}\,(\Psi_0|\bar{\mathbold{\tau}}(E_0)|\Psi_0)}+\cdots\, .
\en
One may evaluate the denominator in this expression by retaining only
leading contributions to $\bar{\mathbold{\tau}}(E_0)$, given by strong
bubbles with Coulomb ladders,
\eq\label{q_tau_p}
({\bf q}|\bar{\mathbold{\tau}}(z)|{\bf p})=
-c_1-\frac{iM_{\pi^0}}{2\pi}\,\rho^{1/2}(z)\, c_2^2
+c_1^2\,\langle\bar{\bf g}_C(z)\rangle+\cdots\, ,
\en
where $\rho(z)=M_{\pi^0}(z-2M_{\pi^0})$, and
 where the quantity $\langle\bar{\bf g}_C(z)\rangle$ is defined in analogy
with Eq.~(\ref{int-g}) for the case of generic $z$. The explicit
expression for this quantity is given by
\eq\label{gc_all}
&&\langle\bar{\bf g}_C(z)\rangle
=\langle\bar{\bf g}_{0-C}(z)\rangle
+\langle\bar{\bf g}_{1-C}(z)\rangle
+\langle\bar{\bf g}_{n-C}(z)\rangle\, ,
\nonumber\\[2mm]
&&\langle\bar{\bf g}_{0-C}(z)\rangle=\frac{M_{\pi^+}}{4\pi}\,
(M_{\pi^+}(2M_{\pi^+}-z))^{1/2}\, ,
\nonumber\\[2mm]
&&\langle\bar{\bf g}_{1-C}(z)\rangle=\frac{\alpha M_{\pi^+}^2}{8\pi}\,
\biggl(\Lambda(\mu)-1+\ln\,\frac{4M_{\pi^+}(2M_{\pi^+}-z)}{\mu^2}\biggr)\, ,
\nonumber\\[2mm]
&&\langle\bar{\bf g}_{n-C}(z)\rangle=\frac{\alpha M_{\pi^+}^2}{4\pi}\,
\biggl(\Psi(2-\eta)-\Psi(1)-\frac{1+2\eta}{1+\eta}\biggr)\, ,
\en
where $\Psi(x)$ denotes the logarithmic derivative of Gamma-function.
The quantity $\eta$ was defined after Eq.~(\ref{Schwinger-I}), and 
$\Lambda(\mu)$ is given in Eq.~(\ref{def_Lambda}).
 Using the above expressions, it is easily seen that the
width is modified at $O(\delta^{11/2})$ in the presence of the
denominator in Eq.~(\ref{denom}). Consequently, at the accuracy we are
working, one may use
\eq
\Gamma=-2\,{\rm Im}\, z=-2\,{\rm Im}\,
(\Psi_0|\bar{\mathbold{\tau}}(E_0)|\Psi_0)\, .
\en

\vspace*{.3cm}

{\bf 2.} In general, the couplings in the non-relativistic Lagrangian are not
real. Decay processes with an energy release at the hard scale
contribute to the imaginary part of these constants. 
The only possible intermediate states in such diagrams are $n\gamma$
and $\pi^0+n\gamma$. Since the anomaly-induced decay into $\pi^0+\gamma$
cannot proceed from the ground state due to $C$-invariance, the
states with a minimal number of photons are $2\gamma$ and $\pi^0+2\gamma$. 
However, the decay width into two photons is of order 
$\delta^5$~\cite{DIRAC,Hammer} ,
and the decay width into
$\pi^0+2\gamma$ starts, at least, at the same order in $\delta$.
Therefore,  at
order $\delta^{9/2}$ one may assume that all couplings in the 
non-relativistic Lagrangian are real, and that the Hamiltonian
constructed from this Lagrangian is hermitian. 

For a hermitian Hamiltonian, the operator $\bar{\mathbold{\tau}}$ 
obeys the unitarity condition
\eq\label{uni}
\bar{\mathbold{\tau}}(E_0)-\bar{\mathbold{\tau}}^\dagger(E_0)=
-2\pi i\bar{\mathbold{\tau}}(E_0)\,\bar\delta(E_0-{\bf H}_0-{\bf H}_C)\,
\bar{\mathbold{\tau}}^\dagger(E_0)\, ,
\en
where the symbol $\bar\delta$ is defined as follows: in order to evaluate 
the right-hand side of Eq.~(\ref{uni}), one  inserts a complete set of
eigenstates  $({\bf H}_0+{\bf H}_C)|\beta\rangle
=E_\beta|\beta\rangle$, omitting the ground state of the 
of the bound $\pi^+\pi^-$ system. 
It is easy to see that the only allowed states are those
containing either $\pi^+\pi^-+N\gamma,~N\geq 0$ or 
$2\pi^0+N\gamma,~N=2k\geq 0$ scattering states
(the decay into $2\pi^0+$~[odd number 
of photons] from the ground state is forbidden by $C$-invariance). 
The contribution from $\pi^+\pi^-+N\gamma$ vanishes due to lack of
phase space. So we have
\eq
\Gamma=\Gamma_{2\pi^0}+\Gamma_{2\pi^0+2\gamma}+\Gamma_{2\pi^0+4\gamma}
+\cdots\, \quad\quad{\rm (real~couplings)}\, .
\en

It can be easily seen that the decay width into $2\pi^0+2\gamma$
starts at $O(\delta^{11/2})$, and the decays into states containing four
and more photons are even more suppressed. Consequently, 
$\Gamma=\Gamma_{2\pi^0}+O(\delta^5)$.

\vspace*{.3cm}

{\bf 3.} 
From the unitarity condition the following expression for the decay width 
into $2\pi^0$ final state is readily obtained,
\eq\label{Gamma2}
\Gamma_{2\pi^0}=\frac{M_{\pi^0}k_0}{4\pi}\,|T_{2\pi^0}(k_0)|^2\, ,\quad\quad
T_{2\pi^0}(k_0)=\int\frac{d^3{\bf p}}{(2\pi)^3}\,\Psi_0({\bf p})
~_+({\bf p}|\bar{\mathbold{\tau}}(E_0)|{\bf k}_0)_0\, ,
\en
where $k_0=|{\bf k}_0|=\{ M_{\pi^0}(E_0-2M_{\pi^0})\}^{1/2}\sim 
O(\delta^{1/2})$ is the magnitude of the relative 3-momentum of the neutral
pion pair in the final state, and the subscripts $+,~0$ 
distinguish between charged $\pi^+\pi^-$ and neutral $\pi^0\pi^0$ state 
vectors, respectively.

The leading strong contribution in $T_{2\pi^0}(k_0)$ starts at 
$O(\delta^{3/2})$. It is straightforwardly seen from Eqs.~(\ref{q_tau_p}), 
(\ref{gc_all}) and (\ref{Gamma2}) that in order to evaluate
width at $O(\delta^{9/2})$, it suffices to calculate $T_{2\pi^0}(k_0)$ at
$O(\delta^{5/2})$. Here we are  interested in the contributions to this
quantity due transverse photons. The diagrams that may potentially
contribute in the lowest order in $\delta$ are depicted in
Fig.~\ref{fig:SE_vertex} - these are the self-energy 
(Figs.~\ref{fig:SE_vertex}a,b) 
and vertex (Figs.~\ref{fig:SE_vertex}d,e) 
corrections to the lowest-order strong 
four-pion non-derivative vertex, with any number of Coulomb photons.
In addition, there are the diagrams Fig.~\ref{fig:SE_vertex}c that stem from 
the counterterm Lagrangian, Eq.~(\ref{Delta-L})  - 
they are needed to renormalize the self-energy diagrams.
These diagrams are the counterparts of diagrams depicted in
Figs.~\ref{fig:topologies}a,b, apart from the fact that in the latter there
are no Coulomb ladders. One may consider the counterparts for other diagrams
depicted in Fig.~\ref{fig:topologies} and Fig.~\ref{fig:topologies_4} which,
however, are expected to be at least as suppressed in powers of $\delta$
in the bound-state calculations, as the ones displayed in
Fig.~\ref{fig:SE_vertex}. 
Further, aiming to establish the power in $\delta$ where these
diagrams start to contribute, we may omit the Coulomb ladders: 
diagrams with  Coulomb photons cannot be amplified with respect to
the diagrams with no Coulomb photons. In order to prove this, we note that
any additional Coulomb exchange in a given diagram
adds an integration measure $d^3{\bf p}$,
the Coulomb potential $\alpha |{\bf p}-{\bf q}|^{-2}$, and the energy
denominator $(z-\cdots-2M_{\pi^+}-{\bf p}^2/M_{\pi^+})^{-1}$. 
As the momenta scale like $\alpha$ or $\alpha^2$, depending on the topology 
of the diagram, it is seen straightforwardly
that a diagram with an additional Coulomb photon, at worst, contributes to
the same order in $\alpha$ as the original one.

The contribution to the matrix element of $\bar{\mathbold{\tau}}$, 
coming from the diagram in
Fig.~\ref{fig:SE_vertex}a (with no Coulomb photons) is
\eq
~_+({\bf p}|\bar{\mathbold{\tau}}(E_0)|
{\bf k}_0)_0^{\rm Fig.\ref{fig:SE_vertex}a}=
-2c_2\frac{\hat\Sigma(\Omega_0,{\bf p})}{\Omega_0} &=& 
-\frac{c_2e^2{\bf p}^2}{M_{\pi^+}^2}\,
\biggl(\frac{\gamma^2+{\bf p}^2}{M_{\pi^+}}\biggr)^{d-3}\,
\frac{\Gamma(d)\Gamma(2-d)}
{(4\pi)^{d/2}\Gamma(1+\frac{d}{2})}+\cdots\, ,\nonumber\\
\hspace*{3cm}
\Omega_0&=&2M_{\pi^+}+\frac{{\bf p}^2}{M_{\pi^+}}-E_0\, ,
\en
where, in order to be consistent with the matching, we have used again the
threshold expansion. The quantity $\hat\Sigma(\Omega_0,{\bf p}^2)$ in this
expression is the self-energy part introduced in Eq.~(\ref{hat-SE}),
calculated at the off-shell value $\Omega_0$ of the 
parameter $\Omega$ - for this reason, one does not encounter an 
infrared  divergence performing the limit $d\rightarrow 3$.  

The contribution from the diagram Fig.~\ref{fig:SE_vertex}a should be
complemented with the counterterm contribution. In the Hamiltonian formalism,
one has first to use the EOM in order to eliminate the time derivative in the
counterterm Lagrangian~(\ref{Delta-L}). We find
\eq
\Delta{\bf H}=-\frac{e^2f_1c_2}{6\pi^2M_{\pi^+}^2}\,\int d^d{\bf x}\,\biggl\{
\sum_\pm\,(\pi_+^\dagger\stackrel{\leftrightarrow}{\bf D}^2\pi_-^\dagger)
(\pi_0\pi_0)+h.c\biggr\}+\cdots\, .
\en
The contribution from this Hamiltonian to the above matrix reads
\eq
~_+({\bf p}|\Delta{\bf H}|{\bf k}_0)_0=
\frac{2c_2e^2f_1}{3\pi^2M_{\pi^+}^2} {\bf p}^2 . 
\en
As was expected, the contribution 
from the counterterm, Fig.~\ref{fig:SE_vertex}c, cancels the UV divergence 
from the self-energy diagrams Fig.~\ref{fig:SE_vertex}a and 
Fig.~\ref{fig:SE_vertex}b. 

Finally, 
folding this result with the ground-state wave function, and rescaling the
integration momentum as ${\bf p}\rightarrow\gamma{\bf p}$, it is seen that
the total (self-energy + counterterm)
contribution to $T_{2\pi^0}(k_0)$ starts at $O(\delta^{9/2})$, 
and therefore can be neglected. Note that one should not be worried about the
(spurious) UV divergence in the integral over the momentum ${\bf p}$ - these
divergences cancel once all contributions in the given order in $\alpha$ are
summed up~\cite{Lamb}.

The contribution coming from the diagram in Fig.~\ref{fig:SE_vertex}d is
 given by 
\eq
&&~_+({\bf p}|\bar{\mathbold{\tau}}(E_0)|
{\bf k}_0)_0^{\rm Fig.\ref{fig:SE_vertex}d}= 
-\frac{e^2c_2}{M_{\pi^+}^2}\,\int\frac{d^d{\bf q}}{(2\pi)^d}\,
\frac{{\bf p}^2{\bf q}^2-({\bf p}{\bf q})^2}{|{\bf p}-{\bf q}|^3}\,
\frac{1}{E_0-2M_{\pi^+}-{\bf q}^2/M_{\pi^+}}
\nonumber\\[2mm]
&&\times\frac{1}{E_0-2M_{\pi^+}-|{\bf p}-{\bf q}|-
{\bf p}^2/(2M_{\pi^+})-{\bf q}^2/(2M_{\pi^+})}
\nonumber\\[2mm]
&&=\frac{e^2c_2}{M_{\pi^+}^2}\,\int\frac{d^d{\bf q}}{(2\pi)^d}\,
\frac{{\bf p}^2{\bf q}^2-({\bf p}{\bf q})^2}{|{\bf p}-{\bf q}|^4}\,
\frac{1}{E_0-2M_{\pi^+}-{\bf q}^2/M_{\pi^+}}+\cdots\, .
\en

Again, the threshold expansion provides one with a homogeneous
expression. After rescaling the momenta, it is immediately seen that the
contribution to the quantity $T_{2\pi^0}(k_0)$ starts at $O(\delta^{7/2})$ 
and therefore can be neglected. The contribution from the 
diagram~\ref{fig:SE_vertex}d
 is identical  to the one  from~\ref{fig:SE_vertex}e. Additional
diagrams may be treated in an analogous manner - using 
power-counting arguments, we expect them to be even more suppressed.

In summary, we  have seen that - for a large class of diagrams - 
transverse photons contribute neither to the decay width 
at $O(\delta^{9/2})$, nor to the matching condition at threshold.
 Although we have not provided a
mathematical proof, we believe that this result is true in general. 
For this reason, we completely neglect
transverse photons in our non-relativistic theory.
The rest is then straightforward: one eliminates
the Coulomb photons by using EOM, and retains only those terms in the
Lagrangian that contribute to the decay width at $O(\delta^{9/2})$.
In this manner, one arrives   at the Lagrangian displayed in 
Eq.~(\ref{Lagr_full}).

\section{Matching procedures}
\label{app:threshold}
\setcounter{equation}{0}

In this paper, we have matched the amplitudes in the relativistic and
non-relativistic theories at physical space dimension $d=3$. In this 
case, both amplitudes contain singular pieces that behave 
like $|{\bf p}|^{-1}$ and $\ln |{\bf p}|$ in the vicinity of the threshold.
In addition, there is the infrared-divergent Coulomb phase. 
The matching is performed at threshold, for the finite parts of the 
amplitudes that are obtained after removing the Coulomb phase, and 
subtracting the singular pieces.

In the literature~\cite{Manohar}, there are examples of a different 
matching condition, where the matching is performed for the full amplitude at
threshold at $d\neq 3$. The threshold singularities then, in general,
transform into  poles at $d=3$ that cancel at a final stage.
In  this Appendix, we compare these matching conditions 
in two specific examples: 
we consider the two diagrams
depicted in Fig.~\ref{fig:soto}, and their non-relativistic counterparts.
As we will see, the two matching conditions lead to exactly the same
result.

Let us start with the vertex correction depicted in Fig.~\ref{fig:soto}a,
that gives rise to the Coulomb phase, and to the $|{\bf p}|^{-1}$ singular
behavior in the real part of the amplitude. In the non-relativistic theory,
the corresponding vertex integral is given by Eqs.~(\ref{Vc_I}), (\ref{Vc})
and (\ref{tc}). If one instead reverses the order of limiting procedures and
calculates the same integral at ${\bf p}=0$, $d\neq 3$, one finds that
the integral vanishes, $\tilde V_c({\bf p},2w({\bf p}))=0$ 
(the symbol ``tilde'' is used to distinguish the quantities 
calculated by using this particular
sequence of limiting procedures). Both the Coulomb phase and the
$|{\bf p}|^{-1}$ singularity disappear when this prescription is used. 

Let us now turn to the same diagram in the relativistic theory.
The infrared-singular contribution at threshold is contained in the function
$G_{+-\gamma}(s)$ defined by 
\eq\label{Gpmg}
G_{+-\gamma}(s)&=&\int\frac{d^D q}{(2\pi)^Di}\,
\frac{1}{q^2(q^2-2qp_1)(q^2+2qp_2)}=
\nonumber\\[2mm]
&-&\frac{2}{s\sigma}\,\biggl\{\ln\frac{1-\sigma}{1+\sigma}+
i\pi\biggr\}\biggl[\lambda_{IR}+\frac{1}{32\pi^2}\,\biggl(
\ln\frac{M_{\pi^+}^2}{\mu^2}+1\biggr)\biggr]
\nonumber\\[2mm]
&-&\frac{1}{32\pi^2s\sigma}\,\biggl(
2\pi i\,\ln\frac{4\sigma^2}{1-\sigma^2}-
4\, {\rm Li}\biggl(\frac{1-\sigma}{1+\sigma}\biggr)+2\pi^2
-\ln^2\frac{1+\sigma}{1-\sigma}\biggr)\, , 
\en
where $p_1^2=p_2^2=M_{\pi^+}^2,~s=(p_1+p_2)^2>4M_{\pi^+}^2,~\sigma=(1-4M_{\pi^+}^2/s)^{1/2}$, and
\eq\label{Li}
{\rm Li}(x)={\rm Li}_2(1-x)=\int_1^x\frac{dt\ln t}{1-t}\, .
\en
This function was considered before in Ref.~\cite{Knecht}, 
where the authors had introduced a  photon mass to tame the
infrared divergences. Here, we use dimensional regularization.
[The definition of ${\rm Li}_2(x)$ in Ref.~\cite{Knecht} 
has to be changed according to Eq.~(\ref{Li})\footnote{
We thank M. Knecht for correspondence on this point.}.]
 
The infrared-divergent quantity $\lambda_{IR}$ is defined by
\eq\label{lambda}
\lambda_{IR}=\frac{\mu^{D-4}}{16\pi^2}\,
\biggl(\frac{1}{D-4}-\frac{1}{2}(\Gamma'(1)+\ln 4\pi+1)\biggr)\, .
\en

The rest of the diagram is infrared-finite at threshold. 
In the vicinity of the threshold,
\eq\label{Gpmg-thr}
e^2G_{+-\gamma}(s)\rightarrow\frac{1}{4M_{\pi^+}^2}\,\biggl\{
-\frac{\pi\alpha M_{\pi^+}}{4|{\bf p}|}-i
\alpha\theta_c+16\pi\alpha\lambda_{IR}
+\frac{\alpha}{2\pi}\,\biggl(\ln\,\frac{M_{\pi^+}^2}{\mu^2}+3\biggr)\biggr\}
\, .
\en
If we reverse the sequence of limiting procedures, we find
\eq\label{t-Gpmg-thr}
e^2\tilde G_{+-\gamma}(s)\rightarrow\frac{1}{4M_{\pi^+}^2}\,\biggl\{
16\pi\alpha\lambda_{IR}
+\frac{\alpha}{2\pi}\,\biggl(\ln\,\frac{M_{\pi^+}^2}{\mu^2}+3\biggr)\biggr\}
\, .
\en
This procedure therefore again amounts to dropping the Coulomb phase
and the singular piece that behaves like $|{\bf p}|^{-1}$. 
The matching condition is not altered since for the particular 
combination of $V_c$ and
$G_{+-\gamma}$ that appears in this condition,
\eq\label{matching-V}
V_c - 4M_{\pi^+}^2G_{+-\gamma}
=\tilde V_c - 4M_{\pi^+}^2\tilde G_{+-\gamma}
=-4\lambda_{IR}-\frac{1}{8\pi^2}\,
\biggl(\ln\,\frac{M_{\pi^+}^2}{\mu^2}+3\biggr)\, .
\en

Let us now consider the diagram depicted in Fig.~\ref{fig:soto}b that leads 
to a logarithmic singularity at threshold. The corresponding loop integral
in the non-relativistic theory is given by Eqs.~(\ref{Bc_I}) and
(\ref{def_Lambda}). If one  reverses the order of limiting procedures,
one finds  $\tilde B_c(2w({\bf p}))=0$. 

In the relativistic theory, it suffices to consider the scalar integral
\eq\label{master-I}
R_c(P)= - e^2\,\int\frac{d^Dl}{(2\pi)^Di}\,\frac{d^Dq}{(2\pi)^Di}D_{\pi^+}(l)
D_{\pi^+}(P-l)D_{\pi^+}(q)D_{\pi^+}(P-q)D(l-q)\, ,
\en
with $D_{\pi^+}(q)=(M_{\pi^+}^2-q^2)^{-1}$, $D(q)=(q^2)^{-1}$. One need not
consider  diagrams with derivative couplings: those can be
expressed through the integral $R_c(P)$ and through integrals that are
 suppressed at threshold as compared to $R_c(P)$, or are infrared-finite.

The explicit expression for $R_c(P)$ is given in Ref.~\cite{Tarasov},
\eq\label{Oleg}
R_c(P)=e^2\frac{\Gamma^2(1+\varepsilon)(M_{\pi^+}^2)^{-1-2\varepsilon}}
{(4\pi)^D\varepsilon(1-2\varepsilon)}\,\biggl\{
\biggl(1-\frac{P^2}{4M_{\pi^+}^2}\biggr)G_4^2
-\frac{H_4}{(1+2\varepsilon)(1-\varepsilon)}\biggr\}\, ,
\en
where $D=4-2\varepsilon$, and
\eq\label{hyper}
G_4=~_2F_1\,\bigl(1,1+\varepsilon;\frac{3}{2};
\frac{P^2}{4M_{\pi^+}^2}\,\bigr)\, ,\quad\quad
H_4=~_3F_2\,\bigl(1,1+\varepsilon,1+2\varepsilon;
\frac{3}{2}+\varepsilon,2-\varepsilon;
\frac{P^2}{4M_{\pi^+}^2}\,\bigr)\, ,
\en
where, in difference with Ref.~\cite{Tarasov}, $P^2$ is defined 
in Minkowski space.

At $D=4$, the integral near threshold in the CM frame 
(${\bf P}={\bf 0}$) 
behaves as~\cite{Beneke}
\eq\label{1-limit}
&&
R_c(P)= - \frac{\alpha}{128\pi M_{\pi^+}^2} 
\biggl(2\ln\frac{2|{\bf p}|}{M_{\pi^+}}
+\ln 2+\frac{21\zeta(3)}{2\pi^2}-i\pi\biggr)+
O(|{\bf p}|)\, ,\quad P^\mu=(2\sqrt{M_{\pi^+}^2+{\bf p}^2},{\bf 0})\, .
\nonumber\\
&&
\en
If we first set ${\bf p}=0$ and then consider the limit $D\rightarrow 4$,
we find
\eq\label{2-limit}
\tilde R_c(P)= - \frac{\alpha}{128\pi M_{\pi^+}^2}
\,\biggl(-\Lambda(\mu)
-\ln\frac{M_{\pi^+}^2}{\mu^2}+1+\ln 2+\frac{21\zeta(3)}{2\pi^2}\biggr)\, ,
\quad P^\mu=(2M_{\pi^+},{\bf 0})\, ,
\en
with $\Lambda(\mu)$ given in (\ref{def_Lambda}).
 For the combination that appears in the
matching condition, we have
\eq\label{matching-I}
B_c-16M_{\pi^+}^4R_c
=\tilde B_c-16M_{\pi^+}^4\tilde R_c 
=\frac{\alpha M_{\pi^+}^2}{8\pi}
\,\biggl(-\Lambda(\mu)
-\ln\frac{M_{\pi^+}^2}{\mu^2}+1+\ln 2+\frac{21\zeta(3)}{2\pi^2}\biggr)\, . 
\en
Consequently, the matching condition remains unaffected by the interchange
of the limiting procedures.

To conclude, for these two diagrams
we have checked that the result of
matching is the same for the two prescriptions.
We expect that this conclusion holds  in the general case as well.

\section{The mapping $SU(3)\times SU(3)\rightarrow SU(2)\times SU(2)$}
\label{app:mapping}\setcounter{equation}{0}

In order to perform the mapping $SU(2)\times SU(2)\rightarrow 
SU(3)\times SU(3)$ for the constants $k_i$ and $K_i$,
 we evaluate the neutral pion mass and the 
amplitude $\pi^+\pi^-\rightarrow\pi^0\pi^0$
in the $SU(3)\times SU(3)$ framework, expand the result in powers of 
$\hat{m}/m_s$ and compare it with its $SU(2)\times SU(2)$ analogue.

From the expressions for the neutral pion mass, we find the relation
 \eq
\label{M-K-k-2}
&&10 k_1^r+10 k_2^r -18 k_3^r +9 k_4^r -10 k_5^r -10 k_6^r -2 k_7^r
\nonumber\\[2mm]
&=&
12 K_1^r+12 K_2^r-18 K_3^r+9 K_4^r
+10 K_5^r+10 K_6^r-12 K_7^r-12 K_8^r-10 K_9^r-10 K_{10}^r\, .
\en
Matching the coefficients of $s$ and $\hat{m}$  in the 
$\pi^+\pi^-\rightarrow\pi^0\pi^0$ amplitudes, we find
\eq
\label{K-k-1}
&&
10 k_1^r
-8 k_2^r
+18 k_3^r
-9 k_4^r
=
 12 K_1^r
-6 K_2^r
+18 K_3^r
-9 K_4^r
+10 K_5^r
-8 K_6^r\, ,
\\[2mm]
\mbox{and}\nonumber \\[2mm]
\label{K-k-2}
&&
20 k_1^r
-16 k_2^r
+18 k_3^r
-9 k_4^r
-10 k_5^r
+26 k_6^r
-2 k_7^r
+36 k_8^r
\nonumber\\[2mm]
&=&24 K_1^r
-12 K_2^r
+18 K_3^r
-9 K_4^r
+20 K_5^r
-16 K_6^r
-12 K_7^r
+24 K_8^r
-10 K_9^r
\nonumber\\[2mm]
&&+26 K_{10}^r
+36 K_{11}^r -144 Z_0 L_4^r
-72 Z_0 L_5^r \, .
\en
Here, $Z_0$ is the $SU(3)\times SU(3)$ analogue of the coupling $Z$. 
In the order of the quark mass expansion
considered here, we may identify $Z_0$ with $Z$.
Combining the relations~(\ref{M-K-k-2}), (\ref{K-k-1}) and
(\ref{K-k-2}), we obtain Eqs. (\ref{kK}) and (\ref{PK_i}).

The coupling $K_i^r$ can be expressed~\cite{Moussallam} as a
convolution of a QCD correlation function with the photon propagator, plus a
contribution from the QED counterterms. We have 
checked that $P(k_i)$ in (\ref{PK_i}) is independent of the 
QCD scale $\mu_0$ that must be
introduced in the QCD Lagrangian after taking into account electromagnetic
effects~\cite{Moussallam,Bijnens}.

\newpage

\begin{center}
{\bf FIGURE CAPTIONS}
\end{center}

\begin{figure}[h]
\epsfxsize=14cm
\epsfysize=4cm
\vspace*{1.cm}\begin{picture}(10,15) \end{picture}
\epsffile{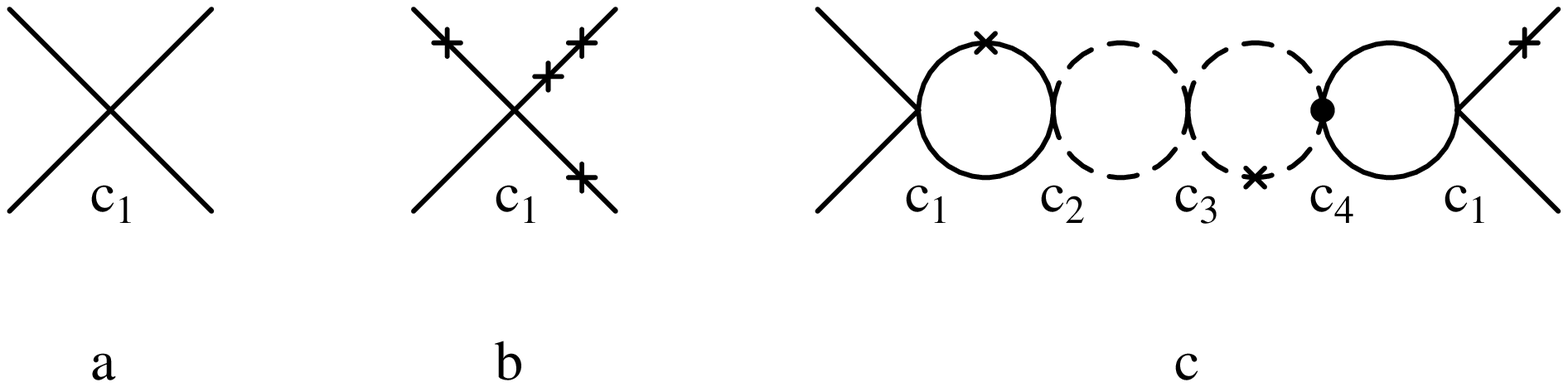}\vspace*{.4cm}
\caption{}\label{fig:strongloops}
\end{figure}

\noindent {\bf FIG.~\ref{fig:strongloops}.}
Examples of diagrams generated by the Lagrangian~(\ref{Lagr_full}) at
$e=0$. Solid (dashed) lines correspond to charged (neutral) pions,
 crosses denote mass insertions, and the filled circle  stands for 
 a higher-order derivative vertex.

\vspace*{.3cm}

\begin{figure}[h]
\epsfxsize=14cm
\epsfysize=4cm
\vspace*{1.cm}\begin{picture}(10,15) \end{picture}
\epsffile{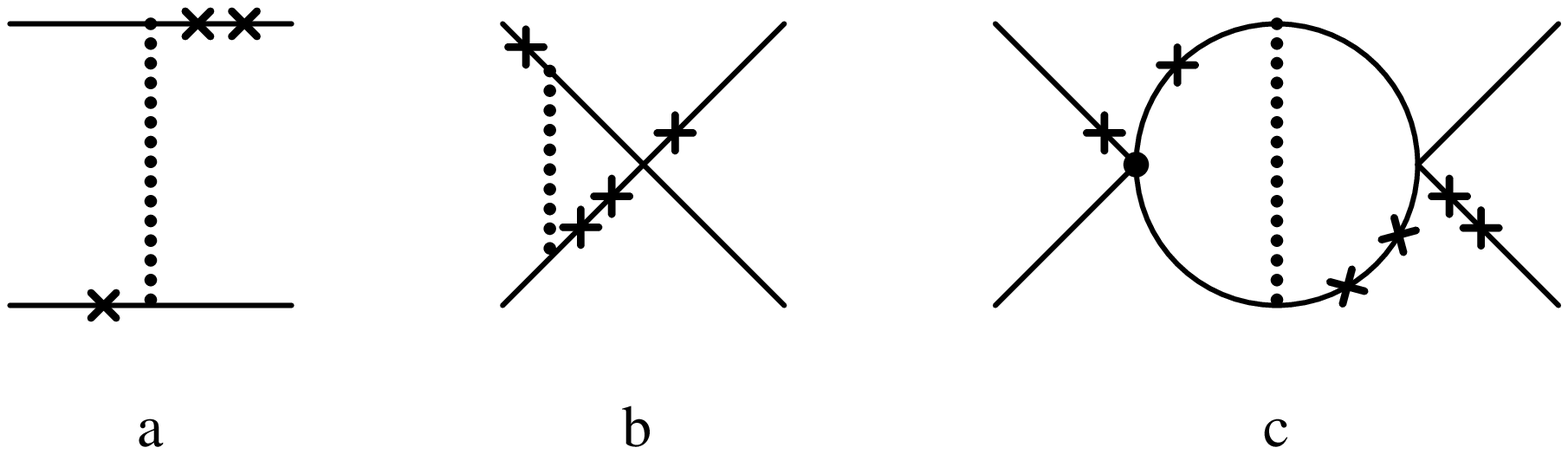}\vspace*{.4cm}
\caption{}\label{fig:Coulomb}
\end{figure}

\noindent {\bf FIG.~\ref{fig:Coulomb}.}
Building blocks for the scattering amplitude, including Coulomb interactions
at order $e^2$. Dotted lines denote the exchange of a Coulomb photon.

\vspace*{.3cm}

\begin{figure}[h]
\epsfxsize=14cm
\epsfysize=6cm
\vspace*{1.cm}\begin{picture}(10,15) \end{picture}
\epsffile{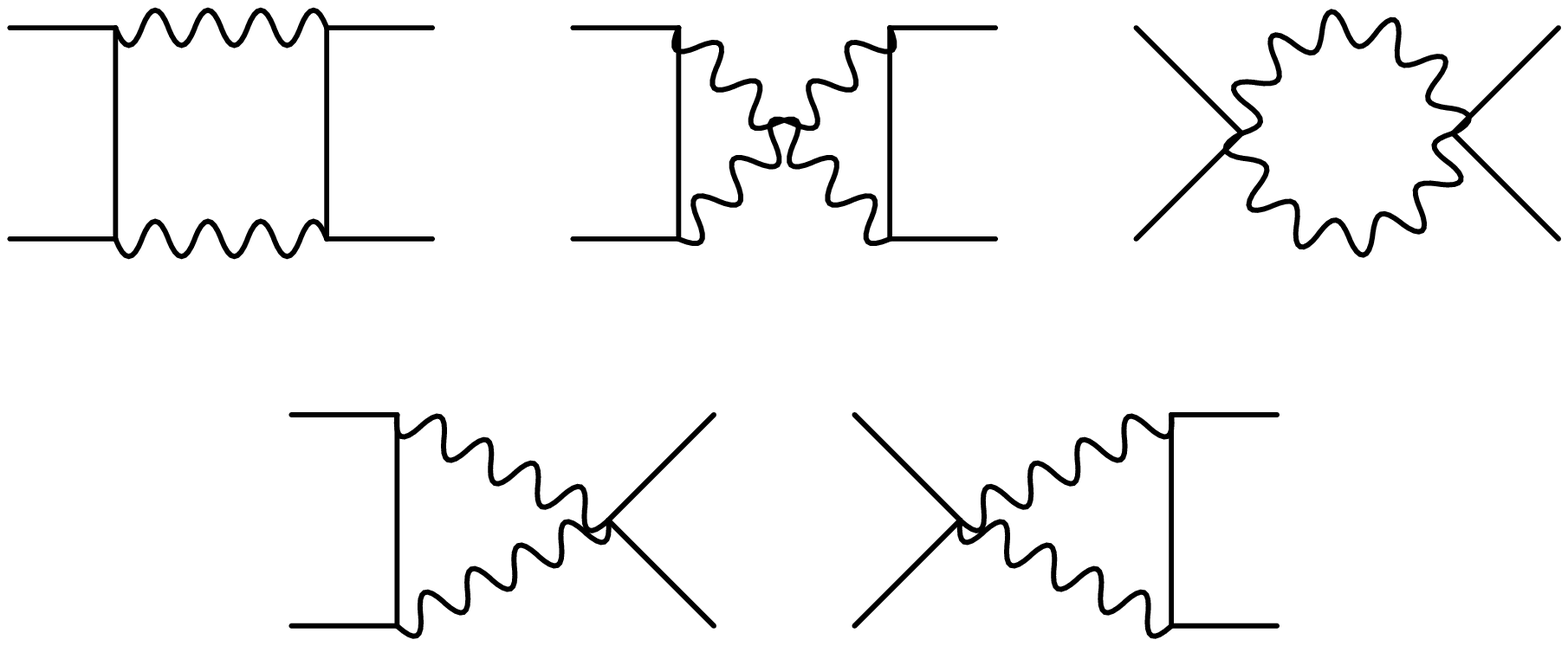}\vspace*{.4cm}
\caption{}\label{fig:2-gamma}
\end{figure}

\noindent {\bf FIG.~\ref{fig:2-gamma}.}
Diagrams contributing to the decay of the $\pi^+\pi^-$ atom into 
two photons in the relativistic theory.

\vspace*{.3cm}

\begin{figure}[h]
\epsfxsize=12cm
\epsfysize=4cm
\vspace*{1.cm}\hspace*{1.cm}\begin{picture}(10,15) \end{picture}
\epsffile{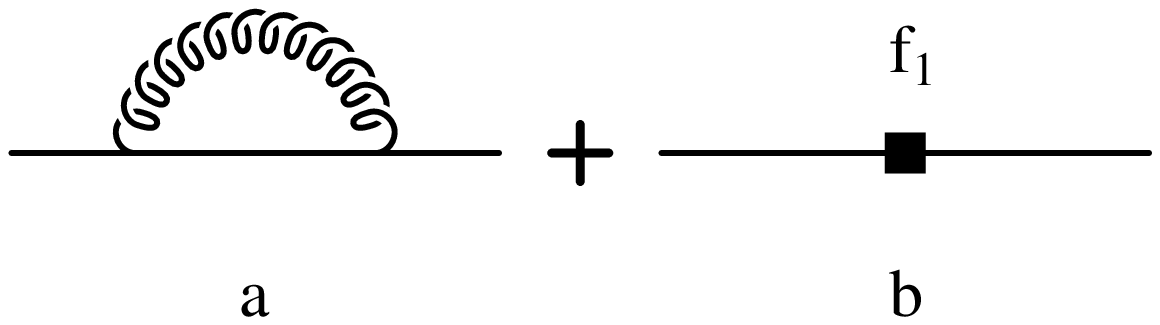}\vspace*{.4cm}
\caption{}\label{fig:SE}
\end{figure}

\noindent {\bf FIG.~\ref{fig:SE}.}
Self-energy of the charged pion at $O(e^2)$. The twisted line denotes
 a transverse photon. The counterterm diagram
b) stems from the Lagrangian~(\ref{Delta-L}).

\vspace*{.3cm}

\begin{figure}[h]
\epsfxsize=14cm
\epsfysize=13cm
\vspace*{1.cm}\begin{picture}(10,15) \end{picture}
\epsffile{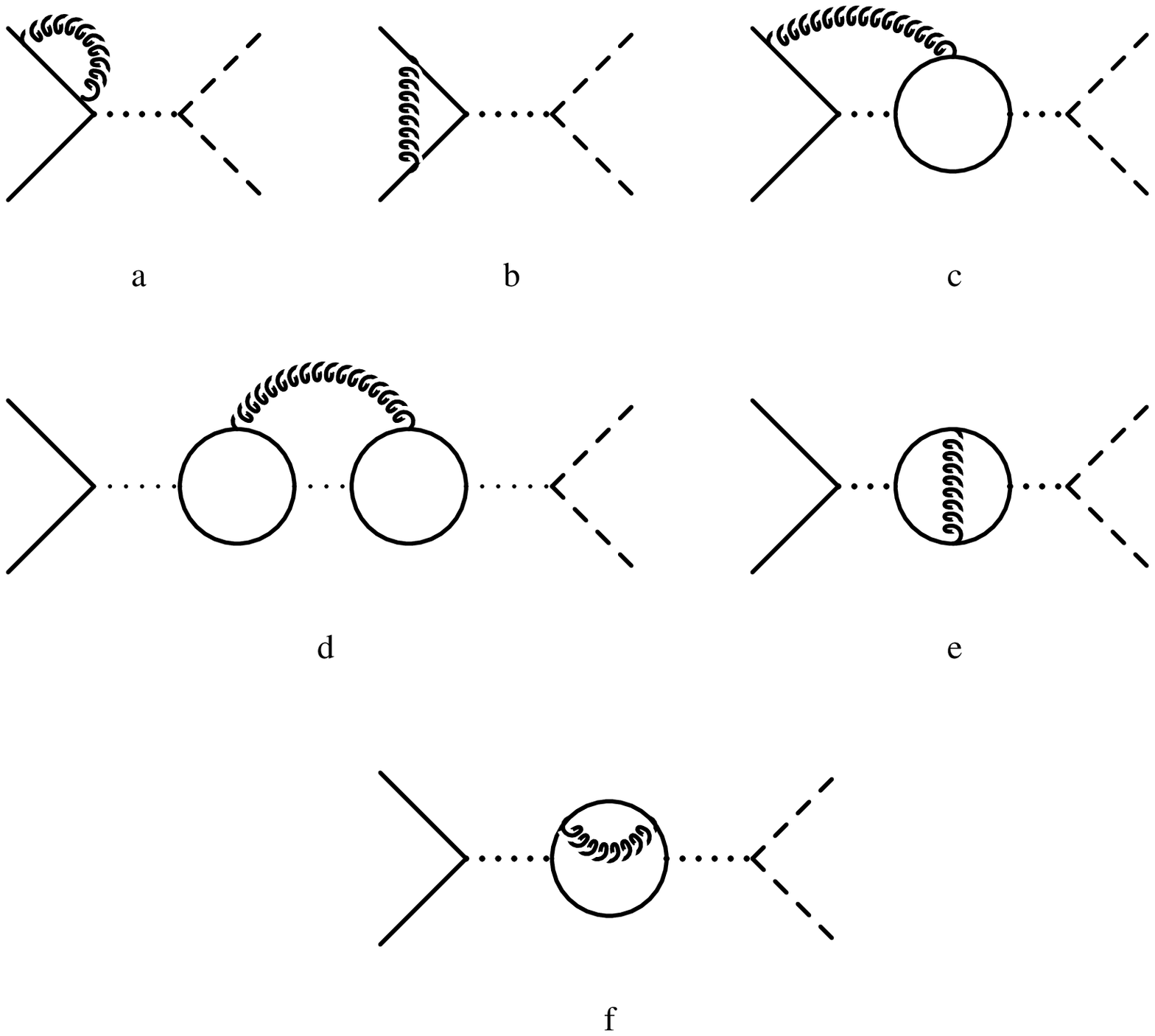}\vspace*{.4cm}
\caption{}\label{fig:topologies}
\end{figure}

\noindent {\bf FIG.~\ref{fig:topologies}.}
Radiative corrections to the scattering amplitude
$\pi^+\pi^-\rightarrow\pi^0\pi^0$, minimal couplings. The twisted lines
 denote transverse photons. Ellipses stand for
any number of charged and neutral pion loops. Mass insertions are not
displayed. 
 
\vspace*{.3cm}

\begin{figure}[h]
\epsfxsize=14cm
\epsfysize=3.3cm
\vspace*{1.cm}\begin{picture}(10,15) \end{picture}
\epsffile{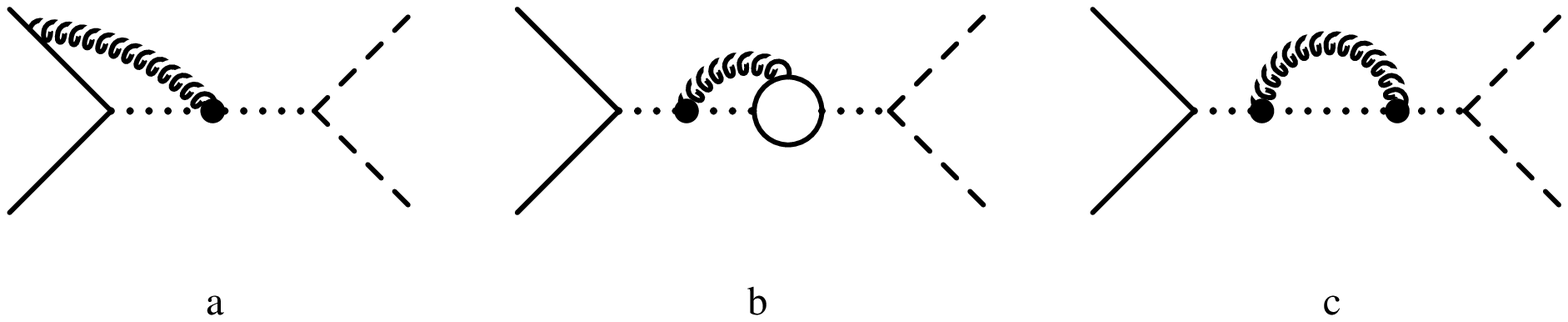}\vspace*{.4cm}
\caption{}\label{fig:topologies_4}
\end{figure}

\noindent {\bf FIG.~\ref{fig:topologies_4}.}
The same as in Fig.~\ref{fig:topologies}, but with at least one vertex
 describing the coupling of a transverse photon to four 
pions (filled circles). 
Ellipses stand for any number of strong loops. 

\vspace*{.3cm}

\begin{figure}[h]
\epsfxsize=14cm
\epsfysize=4cm
\vspace*{1.cm}\begin{picture}(10,15) \end{picture}
\epsffile{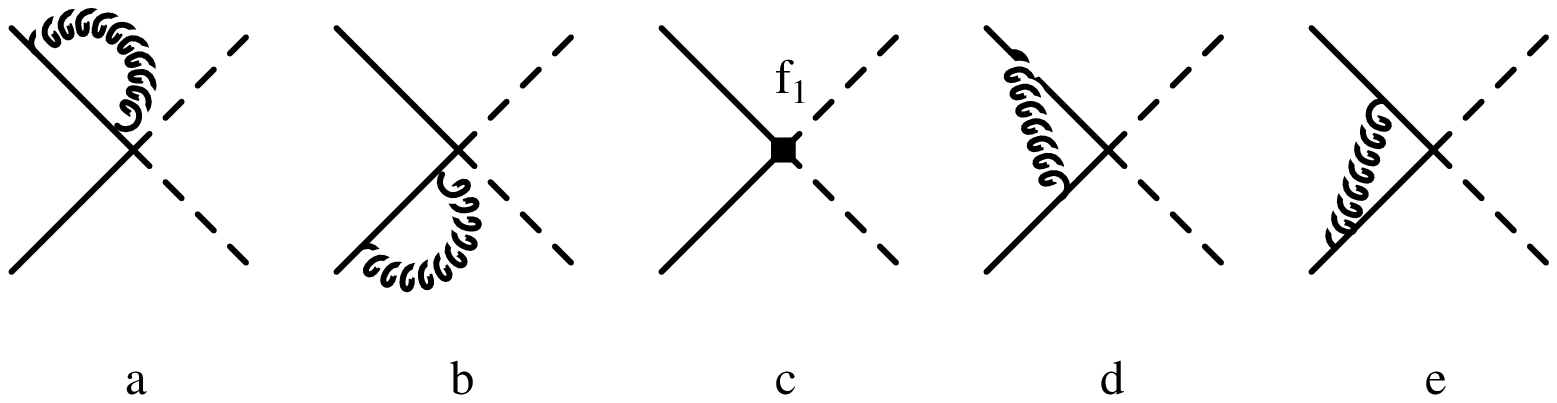}\vspace*{.4cm}
\caption{}\label{fig:SE_vertex}
\end{figure}

\noindent {\bf FIG.~\ref{fig:SE_vertex}.}
Self-energy and vertex corrections  to 
the decay width of the $\pi^+\pi^-$ atom: 
(a,b) self-energy corrections, 
(c) contribution of the counterterm,
(d,e) vertex corrections. The twisted lines stand for transverse photons.

\vspace*{.3cm}

\begin{figure}[h]
\epsfxsize=9cm
\epsfysize=4cm
\vspace*{1.cm}\hspace*{3.cm}\begin{picture}(10,15) \end{picture}
\epsffile{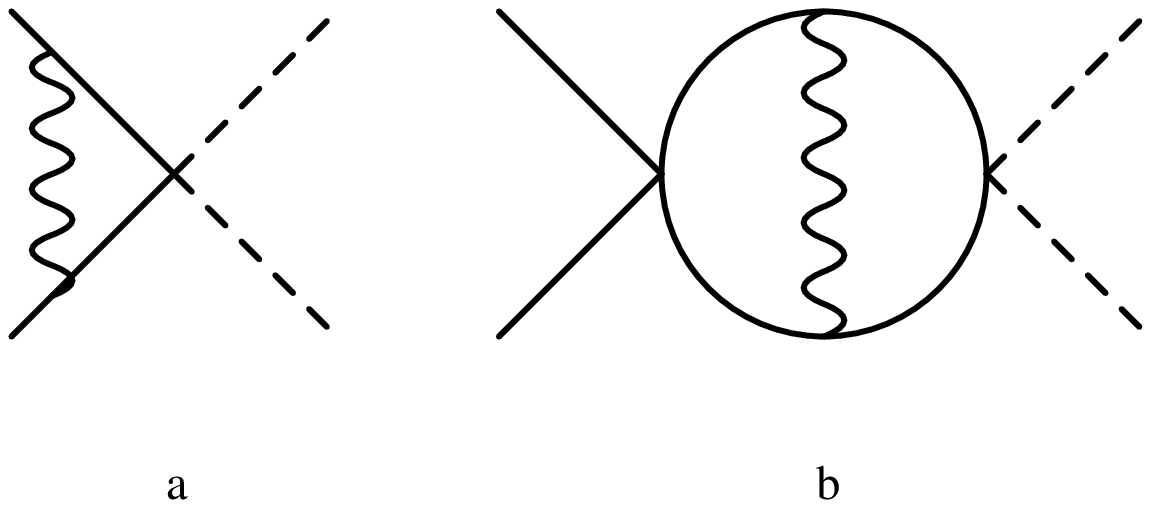}\vspace*{.4cm}
\caption{}\label{fig:soto}
\end{figure}

\noindent {\bf FIG.~\ref{fig:soto}.}
Diagrams that generate a  singular behavior of the
$\pi^+\pi^-\rightarrow\pi^0\pi^0$ scattering amplitude at threshold in the
relativistic theory:
(a) vertex correction,
(b) internal exchange of the photon.

\end{document}